\documentclass[11pt,a4paper]{article}

\usepackage{amssymb}
\usepackage{amsfonts}
\usepackage{amsmath}
\usepackage{latexsym,epsfig}
\usepackage{graphicx} %To include graphics
\usepackage[usenames,dvipsnames]{color} %To use color
\usepackage[hyphens]{url} %To use links. Option 'hyphens' is good for breaking urls.
\usepackage{marvosym} % for the symbol of the cross, and others
\usepackage{soul} %To use highlighting of the text
\usepackage{array} %To use arrays
\usepackage{verbatim}
\usepackage{hyperref}
\usepackage[hang,small,bf]{caption}
\usepackage{url}
\usepackage{enumerate} %To choose enumerator symbols
\newcommand{\LinkedIn}{\textsc{LinkedIn}}

\newcommand{\Twitter}{\textsc{Twitter}}

\newcommand{\ResearchGate}{\textsc{ResearchGate}}
\newcommand{\Academia}{\textsc{Academia}}

\newcommand{\ArnetMiner}{\textsc{ArnetMiner}}
\newcommand{\Egomnia}{\textsc{Egomnia}}
\newcommand{\Google}{\textsc{Google}}
\newcommand{\MathStackExchange}{\textsc{MathStackExchange}}
\newcommand{\MathOverflow}{\textsc{MathOverflow}}
\newcommand{\PageRank}{\textsc{PageRank}}
\newcommand{\TwitterRank}{\textsc{TwitterRank}}
\newcommand{\Hits}{\textsc{Hits}}
\newcommand{\Salsa}{\textsc{Salsa}}

\newtheorem{proposition}{Proposition}%[section]
\title{Endorsement Deduction and Ranking \\ in Social Networks}
\author{Hebert P\'erez-Ros\'es$^{1, 2}$ \thanks{Corresponding author -- E-mail: hebert.perez@matematica.udl.cat -- Office: +34 973 702 781}, \; Francesc Seb\'e$^{1}$ \\
Josep Maria Rib\'o$^{3}$ \Cross  \\ \\
$\phantom{0}^{1}$\emph{Dept. of Mathematics, Universitat de Lleida, Spain} \\
\vspace{2mm} 
$\phantom{0}^{2}$\emph{Conjoint Fellow, University of Newcastle, Australia} \\
\vspace{2mm} 
$\phantom{0}^{3}$\emph{Department of Computer Science and Industrial Engineering,} \\ \emph{Universitat de Lleida, Spain} }
\begin{document}
\maketitle

%%--------------------------------%%
%%            ABSTRACT            %%
%%--------------------------------%%
\vskip 5mm
% \clearpage
\thispagestyle{empty}

\begin{abstract}
Some social networks, such as \LinkedIn \ and \ResearchGate, allow user endorsements for specific skills. In this way, for each skill we get a directed graph where the nodes correspond to users' profiles and the arcs represent endorsement relations. From the number and quality of the endorsements received, an authority score can be assigned to each profile. In this paper we propose an authority score computation method that takes into account the relations existing among different skills. Our method is based on enriching the information contained in the digraph of endorsements corresponding to a specific skill, and then applying a ranking method admitting weighted digraphs, such as \PageRank. We describe the method, and test it on a synthetic network of 1493 nodes, fitted with endorsements. 
% \keywords{Expertise retrieval, directed graphs, social networks, LinkedIn, ResearchGate, PageRank}
\end{abstract}

\vskip 5mm

%%======================================%%
%%            INTRODUCTION              %%
%%======================================%%
\section{Introduction}
\label{sec:intro}

Directed graphs (digraphs) are an appropiate tool for modelling social networks with asymmetric binary relations. For instance, the blogosphere is a social network composed of blogs/bloggers and the directed \lq recommendation\rq \ or \lq follower\rq \ relations among them. Other examples include \lq trust\rq \ statements in recommendation systems (some user states that he$/$she trusts the recommendations given by some other user) and \lq endorsements\rq \ in professional social networks. Additionally, weighted arcs appear in situations where such relations can accomodate some degree of confidence (\lq trust\rq \ or \lq endorsement\rq \ statements could be partial). 

\LinkedIn \ and \ResearchGate \ are two prominent examples of professional social networks implementing the \emph{endorsement} feature. \LinkedIn\footnote{\url{http://www.linkedin.com}} is a wide-scope professional network launched in 2003. More than a decade later it boasts a membership of over 364 million, and it has become an essential tool in professional networking. The \LinkedIn \ endorsement feature, introduced about three years ago,\footnote{More precisely, on September 24, 2012} allows a user to endorse other users for specific skills. 

On the other hand, \ResearchGate\footnote{\url{http://www.researchgate.net}} is a smaller network catering to scientists and academics. It was launched in 2008, and it reached five million members in August, 2014. \ResearchGate \ also introduced an endorsement feature recently.\footnote{On February 7, 2013} From the endorsements shown in an applicant's profile, a potential employer can assess the applicant's skills with a higher level of confidence than say, by just looking at his$/$her CV. 

The two endorsement systems described above are very similar: For each particular skill, the endorsements make up the arcs of a directed graph, whose vertices are the members' profiles. In principle, these endorsement digraphs could be used to compute an authority ranking of the members with respect to each particular skill. This authority ranking may provide a better assessment of a person's profile, and it could become the basis for several social network applications. 

For instance, this authority ranking could be the core element of an eventual tool for finding people who are proficient in a certain skill, very much like a web search engine. It could also find important applications in profile personalization. For example, if a certain user is an expert in some field, say \lq Operations Research\rq, the system can display ads, job openings, and conference announcements related to that field in the user's profile. Finally, we can envisage a world where people could vote on certain decisions via social networks. For example, a community of web developers could decide on the adoption of some particular web standard. In that scenario, we might think about a \emph{weighted voting scheme}, where the weight of each vote is proportional to the person's expertise in that area. 

% In fact, ResearchGate already provides an estimate of scientific reputation, based on the number of publications, number of citations, number of followers, questions, answers, etc.: the RG score. However, it is not possible to get a list of researchers in decreasing order of the RG score, not to mention that the reliability of the RG score is still in debate \cite{lugger12}. 

% To our knowledge, the first social network to compute a ranking of members is the Italian professional social network Egomnia\footnote{\url{http://www.egomnia.com}}, but this ranking is not based on user endorsements. In any case, Egomnia has experienced an explosive growth in Italy, probably due in part to this ranking feature, although it is still confined to the Italian job market.  

Now, people usually have more than one skill, with some of those skills being related. For example, the skill \lq Java\rq \ is a particular case of the skill \lq Programming\rq, which in turn is strongly related with the skill \lq Algorithms\rq. It may well happen that a person is not endorsed for the skill \lq Programming\rq, but he$/$she is endorsed for the skills \lq Java\rq \ and \lq Algorithms\rq. From those endorsements it can be deduced with a fair degree of confidence that the person also possesses the skill \lq Programming\rq. In other words, a person's ranking with respect to the skills \lq Java\rq \ and \lq Algorithms\rq \ affects his$/$her ranking with respect to the skill \lq Programming\rq. 

If the members of a social network were consistent while endorsing their peers, this \lq endorsement with deduction\rq \ would not add anything to simple (i.e. ordinary) endorsement. In this ideal world, if Anna endorses Ben for the skill \lq Java\rq, she would be careful to endorse him for the skill \lq Programming\rq \ as well.\footnote{Some people may argue that knowledge of a programming language does not automatically imply programming skills, but this semantic discussion is out of the scope of this paper.} In practice, however, 

\begin{enumerate}
\item People are not systematic. That is, people do not usually go over all their contacts methodically to endorse, for each contact and alleged skill, all those contacts which, according to their opinion, deserve such endorsement. This may be the source of important omissions in members' profiles. 
\item People are not consistent, for consistency, like method, would require a great effort. In an analysis of a small \LinkedIn \ community consisting of 3250 members we have detected several inconsistencies. For example, there are several users who have been endorsed for some specific programming language, or a combination of programming languages, but have not been endorsed for the skill \lq Programming\rq. Deciding whether there is an inconsistency entails some degree of subjectivism, for inconsistencies ultimately depend on the semantics of the skill names. Nevertheless, we can safely assert that practically 100$\%$ of the profiles sampled by us contained some evident inconsistency or omission. The Appendix lists some of the more significant inconsistencies and omissions encountered, together with a more comprehensive discussion about \LinkedIn's endorsement mechanism. 
\item Skills lack standardization. In most of these social networks, a set of standard, allowed skills has not been defined. As a result, many related skills (in many cases, almost synonyms) may come up in different profiles of the social network. Consider, for example, skills such as \lq recruiting\rq, \lq recrutiments\rq, \lq IT recruiting\rq, \lq internet recruiting\rq, \lq college recruiting\rq, \lq student recruiting\rq, \lq graduate recruiting\rq \ .... which are, all of them, common in \LinkedIn \ profiles. It may well happen that an expert in \lq recruiting\rq \ has not even assigned to him/herself that specific skill, but a related one such as \lq recruitments\rq, which would hide him/her as an expert in the \lq recruiting\rq \ skill.
\end{enumerate}

Endorsement with deduction may help address those problems, and thus provide a better assessment of a person's skills. More precisely, we propose an algorithm that enriches the digraph of endorsements associated to a particular skill with new weighted arcs, taking into account the correlations between that \lq target\rq \ skill and the other ones. Once this has been done, it is possible to apply different ranking algorithms to this enhanced digraph with the purpose of obtaining a ranking of the social network members concerning that specific skill.

% Moreover, the way that LinkedIn implements the endorsement feature adds to the inconsistency. Sometimes, when Anna wants to endorse Ben, she is presented with a group of skills that are bundled together, so that a Anna only has to press a button to endorse Bob for all the skills in the group. We are not familiar with the bundling algorithm, but it is quite clear that the skills are not chosen by their mutual correlation. 

\subsection{Related work}

This research can be inscribed into the discipline of \emph{expertise retrieval}, a sub-field of information retrieval \cite{Bal12}. There are two main problems in expertise retrieval: 

\begin{enumerate}
\item Expert finding: Attempts to answer the question \lq\lq \emph{Who are the experts on topic $X$?}\rq\rq. In our approach, this question is answered by taking all the network members who are within a certain percentile of the ranking for topic $X$. % or equivalently, who have a RageRank value above a certain pre-defined threshold for skill $X$. 
\item Expert profiling: Addresses the question \lq\lq \emph{Which skills does person $Y$ possess?}\rq\rq. We could answer this question by computing the rankings with respect to all the skills claimed by person $Y,$ and taking those skills for which $Y$ has fallen within the pre-defined percentile mentioned above. %, or for which $Y$ has obtained a \PageRank \ value above the pre-defined threshold. 
\end{enumerate}   

Traditionally, these problems above have been solved via document mining, i.e. by looking for the papers on topic $X$ written by person $Y,$ combined with centrality or bibliographic measures, such as the H-index and the G-index, in order to assess the expert's relative influence (e.g. \cite{Li13}). This is also the approach followed by \ArnetMiner\footnote{\url{http://www.arnetminer.org}}, a popular web-based platform for expertise retrieval \cite{arnetminer}. 

Despite their unquestionable usefulness, systems based on document mining, like \ArnetMiner, face formidable challenges that limit their effectiveness. In addition to the specific challenges mentioned by Hashemi et al. \cite{Hash13}, we could add several problems common to all data mining applications (e.g. name disambiguation). As a small experiment, we have searched for some known names in \ArnetMiner, and we get several profiles corresponding to the same person, one for each different spelling. 

That is one of the reasons why other expertise retrieval models resort to the power of \PageRank \ in certain social networks, such as in the perused scientific citation and scientific collaboration networks (e.g. \cite{Den12,Hash13}). Another interesting example related to \PageRank \ and social networks is \TwitterRank \ \cite{Twi10}, an extension of \PageRank \ that measures the relative influence of \Twitter \ users in a certain topic. Like our own \PageRank \ extension, \TwitterRank \ is topic-specific: the random surfer jumps from one user to an acquaintance following topic-dependent probabilities. However, \TwitterRank \ does not consider any relationships among the different topics. 

To the best of our knowledge, there are no precedents for the use of endorsements in social networks, nor for the use of known relationships among different skills, in the context of expertise retrieval. The closest approach might be perhaps the one in \cite{onto10}, which uses the ACM classification system as an ontology that guides the mining process and expert profiling. Another (very recent) model that uses semantic relationships to increase the effectiveness and efficiency of the search is given in \cite{Lee14}. 

Another related field which has attained a growing interest in the last few years is that of reputation systems, that is, systems intended to rank the agents of a domain based on others' agents reports. Strategies for ranking agents in a reputation system range from a direct ranking by agents (as used in eBay) to more sophisticated approaches (see \cite{Kho} for a survey). One particularly important family of reputation system strategies is that of \PageRank-based algorithms. There are many of such approaches. For instance, \cite{CTX} provides an algorithm based on the so-called Dirichlet \PageRank, which addresses problems such as: (1) Some links in the network may indicate distrust rather than trust, and (2) How to infer a ranking for a node based on the ranking stated for a well-known subnetwork.

% \begin{itemize}
% \item Some links in the network may indicate distrust rather than trust.
% \item How to infer a ranking for a node based on the ranking stated for a well-known subnetwork.
% \end{itemize}

Another example of reputation system (again, based on \PageRank) is \cite{Pujol}. In this case, a modification of the \PageRank \ algorithm is used to create a reputation ranking among the members of an academic community. One remarkable issue of this approach is that the network does not exist explicitly, but it is created ad-hoc from the information harvested from the personal web pages of the members (e.g. a couple of members are connected if they have authored a research article together).

A thorough study of reputation systems is clearly beyond the scope of this article, but in any case, all these scenarios above differ significantly from our application for expertise retrieval with deduction of new endorsements, based on existing endorsements of related skills, and information about the correlation between skills. 

%-----------------------------------------------
\subsection{Contribution and plan of this paper}

This paper focuses on professional social networks allowing user endorsements for particular skills, such as \LinkedIn \ and \ResearchGate. Our main contributions can be summarized as follows:

\begin{enumerate}
\item We introduce \emph{endorsement deduction}: an algorithm to enrich$/$enhance the information contained in the digraph of endorsements corresponding to a specific skill (\lq target\rq \ skill or \lq main\rq \ skill) in a social network. This algorithm adds new weighted arcs (corresponding to other skills) to the digraph of endorsements, according to the correlation of the other skills with the \lq main\rq \ skill. We assume the existence of an \lq ontology\rq \ that specifies the relationships among different skills. % and (2) the endorsements that are contained in the digraphs of endorsements corresponding to those related skills. 
\item After this pre-processing we can apply a ranking algorithm to the enriched endorsement digraph, so as to compute an authority score for each network member with respect to the main skill. In particular, we have used the (weighted) \PageRank \ algorithm for that purpose, but in principle, any ranking method could be used, provided that it admits weighted digraphs (e.g. \Hits \ \cite{whits07}). The reasons why we have chosen \PageRank \ in the first place are explained in Section \ref{sub:context}. The authority score obtained by our method could be useful for searching people having a certain skill, for profile personalization, etc. % Thus, the results of a query might be displayed in decreasing order of authority. 
\item We propose a methodology to validate our algorithm, which does not rely as heavily on the human factor as previous validation methods, or on the availability of private information of the members' profiles. More precisely, we discuss the benefits of endorsement deduction in terms of (1) consistency with the results of simple weighted PageRank, (2) reduction in the number of ties and (3) robustness against spamming. Following this methodology, we test our solution on a synthetic network of 1493 nodes and 2489 edges, similar to \LinkedIn, and fitted with endorsements \cite{synthetic}. 
\end{enumerate}

% We propose a method for computing an authority score for each network member with respect to some skill, that takes into account the correlations among different skills. This authority score could be useful for an eventual tool for searching people having a certain skill. Thus, the results of a query might be displayed in decreasing order of authority. 

% The proposed solution takes the endorsements digraph of some particular skill and adds some (weighted) arcs so as to include information on related skills. We assume the existence of an \lq ontology\rq \ that specifies the relationships among different skills. After that, the (weighted) \PageRank \ algorithm is applied so that each node is assigned a score. We test our solution on a synthetic network of 1493 nodes and 2489 edges, similar to \LinkedIn, and fitted with endorsements \cite{synthetic}.   

The rest of the paper is organized as follows: Section~\ref{sec:preliminaries} provides the essential concepts, terminology and notation that will be used throughout the rest of the paper. It also describes the \PageRank \ algorithm, including the variant for weighted digraphs. After that, our proposal is explained in Section~\ref{sec:deductive} together with a simple example. In Section \ref{sec:comparison} we compare the results obtained by ranking with deduction with those obtained by simple ranking, according to three criteria proposed by ourselves. Finally, in Section \ref{sec:open} we summarize our results, discuss some potential applications, and enumerate some open problems that arise as an immediate consequence of the preceding discussion.   

%%============================%%
%%        NOTATION            %%
%%============================%%
\section{Preliminaries}
\label{sec:preliminaries}

\subsection{Terminology and notation}
\label{sec:notation}

A \emph{directed graph}, or {\em digraph\/} $D=(V,A)$ is a finite nonempty set $V$ of objects called {\it vertices\/} and a set $A$ of ordered pairs of vertices called {\it arcs\/}. The {\it order\/} of $D$ is the cardinality of its set of vertices $V$. If $(u,v)$ is an arc, it is said that $v$ is {\em adjacent from\/} $u$. The set of vertices that are adjacent from a given vertex $u$ is denoted by $N^{+}(u)$ and its cardinality is the {\em out-degree\/} of $u$, $d^{+}(u)$. 

Given a digraph $D=(V,A)$ of order $n$, the adjacency matrix of $D$ is an $n \times n$ matrix $\mathbf{M}=(m_{ij})_{n \times n}$ with $m_{ij}=1$ if $(v_i,v_j) \in A$, and $m_{ij}=0$ otherwise. The sum of all elements in the $i$-th row of $M$ will be denoted $\Sigma m_{i*}$, and it corresponds to $d^{+}(v_i)$. 

A \emph{weighted digraph} is a digraph with (numeric) labels or \emph{weights} attached to its arcs. Given $(u,v)\in A$, $\omega(u,v)$ denotes the weight attached to that arc. In this paper we only consider directed graphs with non-negative weights. The reader is referred to Chartrand and Lesniak \cite{CL} for additional concepts on digraphs. 

%-----------------------------------------------
\subsection{\PageRank \ vector of a digraph}
\label{sec:PageRank} 
 
\PageRank~\cite{Ber05,PBMW98} is a link analysis algorithm that assigns a numerical weighting to the vertices of a directed graph. The weighting assigned to each vertex can be interpreted as a relevance score of that vertex inside the digraph.

The idea behind \PageRank \ is that the relevance of a vertex increases when it is linked from relevant vertices. Given a directed graph $D=(V,A)$ of order $n$, assuming each vertex has at least one outlink, we define the $n \times n$ matrix $\mathbf{P}=(p_{ij})_{n \times n}$ as, 

\begin{equation}
\label{eq:linkmatrix1}
p_{ij}=
\left\{
\begin{array}{cl}
\frac{1}{d^+(v_i)} & \textrm{if } (v_i,v_j) \in A, \\
0 & \textrm{otherwise.} 
\end{array}
\right.
\end{equation}

Those vertices without oulinks are considered as if they had an outlink pointing to each vertex in $D$ (including a loop link pointing to themselves). That is, if $d^+(v_i)=0$ then $p_{ij}=1/n$ for each $j$. Note that $\mathbf{P}$ is a stochastic matrix whose coefficient $p_{ij}$ can be viewed as the probability that a surfer located at vertex $v_i$ jumps to vertex $v_j$, under the assumption that the next movement is taken uniformly at random among the arcs emanating from $v_i$. When the surfer falls into a vertex $v_i$ such that $d^+(v_i)=0$, then he$/$she is able to restart the navigation from any vertex of $D$ uniformly chosen at random. So as to permit this random restart behaviour when the surfer is at any vertex (with a small probability $1-\alpha$), a new matrix $\mathbf{P}_{\alpha}$ is created as, 

\begin{equation}
\label{eq:alpha}
\mathbf{P}_{\alpha} = \alpha \mathbf{P} + (1-\alpha) \frac{1}{n}\mathbf{J}^{(n)},
\end{equation}

where $\mathbf{J}^{(n)}$ denotes the order-$n$ all-ones square matrix.

By construction, $\mathbf{P}_{\alpha}$ is a positive matrix~\cite{M01}, hence, $\mathbf{P}_{\alpha}$ has a unique positive eigenvalue
(whose value is $1$) on the spectral circle. The \PageRank \ {\em vector} is defined to be the (positive) left-hand eigenvector $\mathcal{P} = (p_1,\ldots,p_{n})$ with $\sum_i p_i=1$ (the left-hand Perron vector of $\mathbf{P}_{\alpha}$) associated to this eigenvalue. The probability $\alpha$, known as the {\em damping factor}, is usually chosen to be $\alpha=0.85$. 

The relevance score assigned by \PageRank \ to vertex $v_i$ is $p_i$. This value represents the long-run fraction of time the surfer would spend at vertex $v_i$. 

%-----------------------------------------------
\subsection{\PageRank \ vector of a weighted digraph}

When the input digraph is weighted, the \PageRank \ algorithm is easily adapted so that the probability that the random surfer follows a certain link is proportional to its (positive) weight~\cite{WG04}. This is achieved by slightly modifying the definition, previously given in Eq.~\ref{eq:linkmatrix1}, of matrix ${\bf P}$ so that, 

\begin{equation}
\label{eq:linkmatrix2}
p_{ij}=
\left\{
\begin{array}{cl}
\frac{\omega(v_i,v_j)}{\sum_{v \in N^+(v_i)} \omega(v_i,v)} & \textrm{if } (v_i,v_j) \in A, \\
0 & \textrm{otherwise.} 
\end{array}
\right.
\end{equation}
 
Nodes with no outlinks are treated in the same way as before.

%-----------------------------------------------
\subsection{Personalized \PageRank}
\label{sub:personalized}

\textsc{Personalized PageRank}~\cite{Hav2003} is a variant of \PageRank \ in which, when the surfer performs a random
restart (with probability $1-\alpha$), the vertex it moves to is chosen at random according to
a {\em personalization vector} $\mathbf{v=(v_i)}$ so that $\mathbf{v_i}$ is the probability of restarting navigation from vertex $v_i$.
Now, matrix $\mathbf{P}_{\alpha}$ is computed as,

\begin{equation}
\label{eq:alpha}
\mathbf{P}_{\alpha} = \alpha \mathbf{P} + (1-\alpha) \mathbf{ev}^T,
\end{equation}
with $\mathbf{e}$ denoting the order-$n$ all ones vector. As a result, 
the computation is biased to increase the effect of those vertices
$v_i$ receiving a larger $\mathbf{v_i}$. 

%-----------------------------------------------
\subsection{\PageRank \ in context}
\label{sub:context}

\PageRank \ is actually a variant of spectral ranking, a family of ranking techniques based on eigenvalues and eigenvectors. Vigna \cite{Vigna13} traces the origins of spectral ranking to the 1950's, with \cite{Kat53} and \cite{Wei52}. Afterwards, the method was  rediscovered several times until the 1970's. Other articles which are frequently cited as the original sources of spectral ranking include \cite{Gou67,Bona72,Pin76}. Eventually, the method became widely popular when it was adopted by \Google \ for its search engine. 

The reasons for the popularity of spectral ranking in general, and \Google's \PageRank \ in particular, are, in the first place, their nice mathematical properties. Under some reasonable mathematical assumptions, \PageRank \ produces a unique ranking vector, which reflects very accurately the relative importance of the nodes. Other competing algorithms, such as \Hits \ and \Salsa \ do not guarantee such properties \cite{Fara06}. As we have seen in the previous subsections, \PageRank \ can be adapted to weighted digraphs and supports personalization. Additionally, it can be efficiently approximated \cite{Bar08,Bro06}, and can be computed in a parallel or distributed framework \cite{Kohl06,Sar15}. 

Besides information retrieval, spectral ranking in general, and \PageRank \ in particular, have been applied in social network analysis \cite{Bona72,Twi10,Pe13}, scientometrics \cite{Pin76,Ma08,SJR10,Yan11}, geographic networks \cite{Gou67}, and many other areas with great success. 

Last but not least, \Google's \PageRank \ has withstood the test of public scrutiny, as it has been validated by millions of users for more than fifteen years now.

%%===================================%%
%%     DEDUCTION AND RANKING         %%
%%===================================%%

\section{Endorsement deduction and ranking}
\label{sec:deductive}

Let us consider a professional network in which users can indicate a set of topics they are skilled in. So as to attract attention, some dishonest network members could be tempted to set an over-inflated skill list. The effect of such malicious behaviour is reduced if network members are able to endorse other users for specific skills and the relevance they get depends on the received endorsements. Since
cheating users will rarely be endorsed, their relevance in the network will be kept low. 

In such a social network we get an endorsement digraph for each skill. Our objective is to compute an authority ranking for a particular skill, which is not only based on the endorsement digraph of that particular skill, but also takes into account the endorsement digraphs of other related skills. From now on, the skill for which we want to compute the ranking will be called the \emph{main skill}. 

Let $S = \{ s_0, s_1, \ldots s_\ell \}$ be the set of all possible skills, with $s_0$ being the main skill. Let $D_k = (V, A_k)$ denote the endorsement digraph corresponding to skill $s_k$, and let $\mathbf{M}_{k}$ be its adjacency matrix. 
 
We now define the \emph{skill deduction matrix}  $\mathbf{\Pi} = (\pi_{kt})$ as follows: Given a pair of skills $s_k$ and $s_t$, $\pi_{kt}$ represents the probability that a person skilled in $s_k$ also possesses the skill $s_t$. In other words, from $s_k$ we can infer $s_t$ with a degree of confidence $\pi_{kt}$. By definition, $\pi_{kk} = 1$ for all $k$. In this way, if some user endorses another user for skill $s_k$ but no endorsement is provided for skill $s_t$, we can deduce that an endorsement (for $s_t$) should really be there with probability $\pi_{kt}$. In general, $\mathbf{\Pi}$ will be non-symmetric and sparse, thus it is better represented as a directed graph with weighted arcs. 

Note that $\mathbf{\Pi}$ can be seen as an ontology that also accounts for hierarchies among the topics. For example, \lq Applied Mathematics\rq \ is a sub-category of \lq Mathematics\rq, and this would be reflected in $\mathbf{\Pi}$ as a link with weight $1$, going from \lq Applied Mathematics\rq \ to \lq Mathematics\rq.

Our proposal takes as input the skill deduction matrix $\mathbf{\Pi}$, together with those endorsement digraphs $D_{k}$, with $0 < k \leq \ell$, such that $\pi_{k0} > 0$. Without loss of generality, we will assume that the set of skills related to $s_0$ is $S_0 = \{s_k \;\;|\;\; k\neq 0,\;\; \pi_{k0} > 0\} = \{s_1,\ldots,s_\ell\}$.%\footnote{Without loss of generality we can assume that all skills are related to $s_0$, some of them with probability $\pi_{k0} = 0$.}

The proposed endorsement deduction method constructs a weighted endorsement digraph $D_0^{we}=(V,A_0^{we})$ on skill $s_0$, with weights ranging from $0$ to $1$, considering the endorsements deduced from related skills $\{s_1,\ldots,s_{\ell}\}$. 

\begin{enumerate}
\item First of all, if user $v_i$ directly endorsed $v_j$ for skill $s_0$, that is $(v_i,v_j) \in A_0$, then $D_0^{we}$ has arc $(v_i,v_j)\in A_0^{we}$ with $\omega(v_i,v_j)=1$ (that endorsement receives a maximum confidence level). 

\item If $(v_i, v_j) \notin A_0$ but $(v_i, v_j) \in A_{k}$, for just one $k$, $1 \leq k \leq \ell$, then arc $(v_i, v_j)$ is added to $D_0^{we}$ with weight $\omega(v_i, v_j) = \pi_{k0}$, that is, the arc is assigned a weight that corresponds to the probability that $v_i$ also considers $v_j$ proficient in skill $s_0$, given an existing endorsement for skill $s_k$. 

\item Finally, if $(v_i,v_j)\notin A_0$ but $(v_i,v_j)\in A_{k_1},\ldots,A_{k_\ell}$, then the arc $(v_i,v_j)$ is assigned a weight corresponding to the probability that $v_i$ would endorse $v_j$ for $s_0$ given his$/$her endorsements for $s_{k_1},\ldots,s_{k_l}$.
That is, let ``$(s_{k_i}\rightarrow s_0)$'' denote the event ``endorse for skill $s_0$ given an endorsement for $s_{k_i}$ (its probability is $p(s_{k_i}\rightarrow s_0)=\pi_{k_i,0}$) then $(v_i,v_j)$ is assigned a weight that corresponds to the probability of 
the union event ``$\cup_{k_i \in \{ k_1, \ldots, k_\ell \}} (s_{k_i}\rightarrow s_0)$'', assuming those events are independent.
\end{enumerate}

Next we show how to construct the weighted adjacency matrix of $D_0^{we}$ by iteratively adding deduced information from related skills. Computations are shown in Eqs.~\ref{eq:deductive}. After the $k$-th iteration, matrix $\mathbf{Q}_k$ corresponds to the weighted digraph of skill $s_0$ after having added deduced information from skills $s_1,\ldots,s_k$. The matrix computed after the last iteration $\mathbf{Q}_\ell$ corresponds to the weighted adjacency matrix of digraph $D_0^{we}$. Computations can be carried out as follows,

%----------------------
% \begin{eqnarray}
% \label{eq:deductive}
%   \mathbf{Q}_0 &=& \mathbf{M}_0;  \\
%   \mathbf{Q}_k &=& \mathbf{Q}_{k-1} + \pi_{k0} ((\mathbf{J}^{(n)}-\mathbf{Q}_{k-1}) \circ \mathbf{M}_k), 
%   \; \mbox{ for } k = 1, \ldots, \ell; \nonumber 
% \end{eqnarray}

%\begin{equation}
\begin{subequations} 
\label{eq:deductive}
\begin{align} 
   \mathbf{Q}_0 \; &=& & \mathbf{M}_0  \label{eq:firstcase} \\
   \mathbf{Q}_k \; &=& & \mathbf{Q}_{k-1} + \pi_{k0} ((\mathbf{J}^{(n)}-\mathbf{Q}_{k-1}) \circ \mathbf{M}_k),
   \; \mbox{ for } k = 1, \ldots, \ell, \label{eq:secondcase}
\end{align} 
\end{subequations}
%\end{equation}
%----------------------

where the symbol \lq $\circ$\rq \ represents the Hadamard or elementwise product of matrices. 

Note that Eq. \ref{eq:secondcase} acts on the entries of $\mathbf{Q}_{k-1}$ that are smaller than $1$, and the entries equal to $1$ are left untouched. If some entry $\mathbf{Q}_{k-1}(i,j)$ is zero, and the corresponding entry $\mathbf{M}_k(i,j)$ is non-zero, then $\mathbf{Q}_{k-1}(i,j)$ takes the value of $\mathbf{M}_k(i,j)$, modified by the weight $\pi_{k0}$. This corresponds to the second case above. 

If $\mathbf{Q}_{k-1}(i,j)$ and $\mathbf{M}_k(i,j)$ are both non-zero, then we are in the third case above. To see how it works, let us suppose that some entry $\mathbf{M}_0(i,j)$ is zero, but the corresponding entries $\mathbf{M}_1(i,j), \mathbf{M}_2(i,j), \mathbf{M}_3(i,j), \ldots$, are all equal to $1$. In other words, person $i$ does not endorse person $j$ for the main skill (skill $0$), but does endorse person $j$ for skills $1, 2, 3, \ldots$. In order to simplify the notation we will drop the subscripts $i,j$, and we will refer to $q_k$ as the $(i,j)$-entry of $\mathbf{Q}_{k}$. Applying Equations \ref{eq:deductive}, we get:

\begin{eqnarray*}
q_0 &=& m_0 = 0 \\
q_1 &=& q_0 + \pi_{1,0}(1-q_0) = \pi_{1,0} \\
q_2 &=& q_1 + \pi_{2,0}(1-q_1) = \pi_{1,0} + \pi_{2,0}(1-\pi_{1,0}) \\
    &=& \pi_{1,0} + \pi_{2,0} - \pi_{1,0}\pi_{2,0} \\
q_3 &=&  q_2 + \pi_{3,0}(1-q_2) \\
    &=& \pi_{1,0} + \pi_{2,0} - \pi_{1,0}\pi_{2,0} + \pi_{3,0}(1-(\pi_{1,0} + \pi_{2,0} - \pi_{1,0}\pi_{2,0})) \\
    &=& \pi_{1,0} + \pi_{2,0} + \pi_{3,0} - \pi_{1,0}\pi_{2,0} - \pi_{1,0}\pi_{3,0} - \pi_{2,0}\pi_{3,0} + \pi_{1,0}\pi_{2,0}\pi_{3,0} \\
    &\vdots&  
\end{eqnarray*}

which corresponds to the probabilities of the events $(s_1 \rightarrow s_0)$, $(s_1 \rightarrow s_0) \cup (s_2 \rightarrow s_0)$, $(s_1 \rightarrow s_0) \cup (s_2 \rightarrow s_0) \cup (s_3 \rightarrow s_0)$, and so on.  

Once we have the matrix $\mathbf{Q}_\ell=(q_{ij})_{n\times n}$, we can apply any ranking method that admits weighted digraphs, such as the weighted \PageRank \ algorithm \cite{WG04}. For that purpose we have to construct the normalized weighted link matrix $\mathbf{P}$, as in Eq.~\ref{eq:linkmatrix2}: 

\begin{equation}
\label{eq:pesos}
p_{ij}=
\left\{
\begin{array}{cl}
\frac{q_{ij}}{\Sigma q_{i*}} & \textrm{if } \Sigma q_{i*}>0, \\\\
\frac{1}{n} & \textrm{if } \Sigma q_{i*}=0. 
% 1/n & \textrm{if } \Sigma q_{i*}=0. 
\end{array}
\right.
\end{equation}

Then we compute $\mathbf{P}_{\alpha}$ from $\mathbf{P}$, as in Eq.~\ref{eq:alpha}, and we finally apply the weighted \PageRank \ algorithm on $\mathbf{P}_{\alpha}$. 

%----------------------
\subsection{An example}
%----------------------

\begin{figure}[htbp]
\begin{center}
%	 \centering
	 	\includegraphics[width=0.45\textwidth]{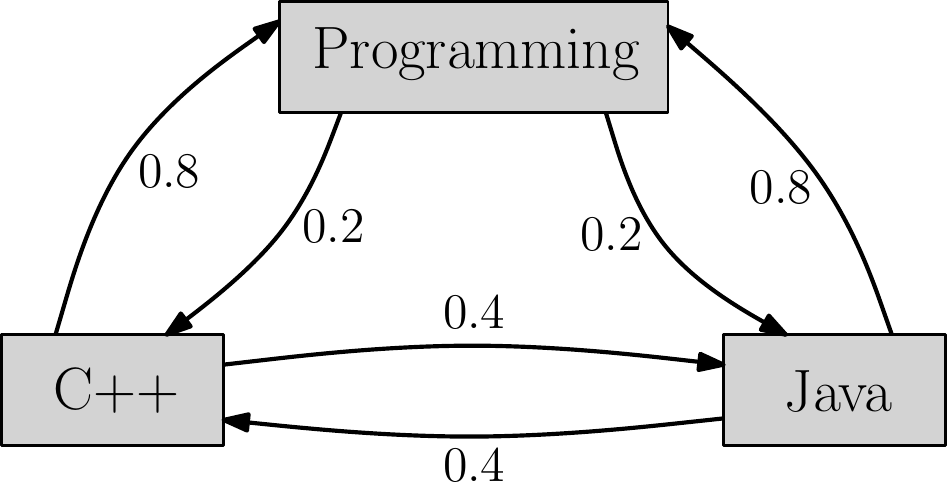}
	 \caption{Directed graph representing a skill deduction matrix $\mathbf{\Pi}$.}
	 \label{fig:skills}
\end{center}	
\end{figure}

As a simple illustration, let us consider a set of three skills: \lq Programming\rq,\ \lq C++\rq\ and \lq Java\rq. The probabilities relating them, depicted in Figure~\ref{fig:skills}, have been chosen arbitrarily, but in practice, they could have been obtained as a result of some statistical analysis. % In this example we will only consider two of the skills, which are colored in grey.  

%--------------------
%--------------------

\begin{figure}[htbp]
\begin{center}
%	 \centering
	 	\includegraphics[width=0.8\textwidth]{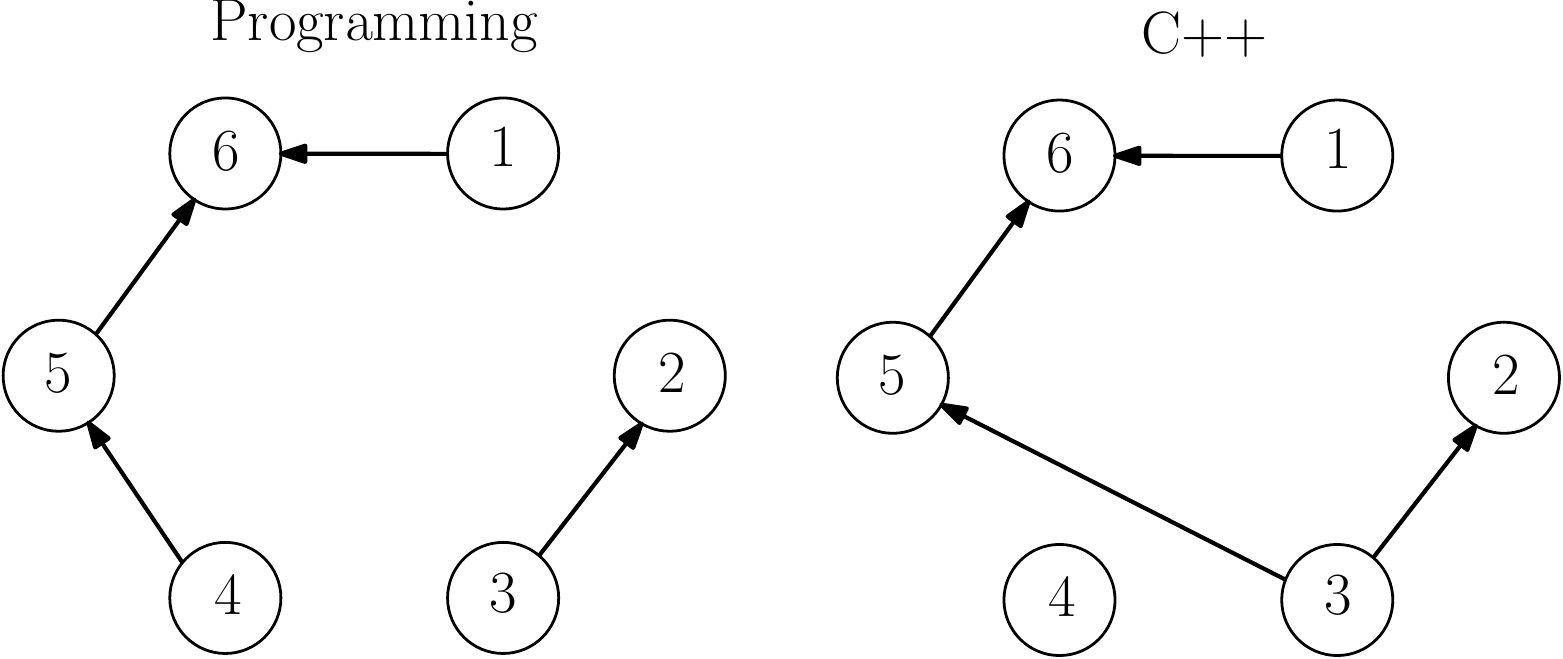}
	 \caption{Endorsements for \lq Programming\rq (left) \ and \lq C++\rq (right).}
	 \label{fig:endorsements1}
\end{center}	
\end{figure}

Let us further assume that we have a community of six individuals, labeled from \lq 1\rq \ to \lq 6\rq. 
Figure~\ref{fig:endorsements1} shows the endorsement digraphs among the community members for the skills \lq Programming\rq \ and \lq C++\rq. 

%--------------------
%--------------------

Let us suppose that the skill \lq Programming\rq \ is our main skill (skill $0$). Thus, $\mathbf{Q}_0 = \mathbf{M}_0$ is the adjacency matrix of the digraph shown in Figure~\ref{fig:endorsements1} (left). If we compute the \PageRank \ for the skill \lq Programming\rq, without considering its relationships with other skills, we get the following scores ($\mathcal{P}(v)$ denotes the \PageRank \ score assigned to vertex $v$): $$\mathcal{P}(1) = \mathcal{P}(3) = \mathcal{P}(4) = 0.0988,\; \mathcal{P}(2) = \mathcal{P}(5) = 0.1828,\; {\rm and}\; \mathcal{P}(6) = 0.3380.$$
 
In other words, on the basis of the endorsements for \lq Programming\rq \ alone, the individuals \lq 2\rq \ and \lq 5\rq \ are tied up, and hence equally ranked. 

\begin{figure}[htbp]
\begin{center}
%	 \centering
 	\includegraphics[width=0.4\textwidth]{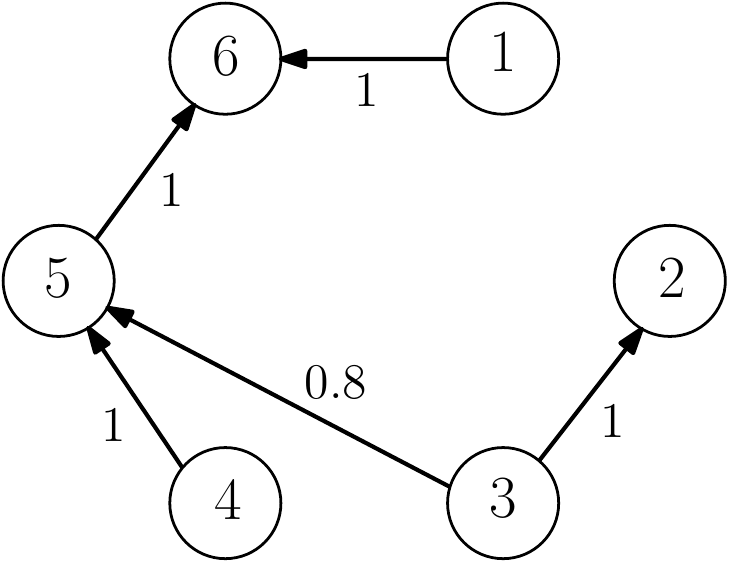}
	 \caption{Endorsements for \lq Programming\rq, with information deduced from \lq C++\rq.}
	 \label{fig:endorsements2}
\end{center}	
\end{figure}

Now we will include the endorsements for \lq C++\rq \ in this analysis (skill $1$). We apply Eq. \ref{eq:deductive} to compute $\mathbf{Q}_1$, as follows:

$$
\mathbf{Q}_1 = \mathbf{Q}_{0} + \pi_{1,0} ((\mathbf{J}^{(6)} - \mathbf{Q}_{0}) \circ \mathbf{M}_1), 
$$

where $\pi_{1,0} = 0.8$, and $\mathbf{M}_1$ is the adjacency matrix of the digraph shown in Figure~\ref{fig:endorsements1} (right). This yields the endorsement digraph depicted in Figure~\ref{fig:endorsements2}. 

% \begin{equation}
%   \label{eq:q1}
%   \mathbf{Q}_1 = \left( \begin{array}{cccccc}
%   0 & 0 & 0 & 0 & 0   & 1 \\
%   0 & 0 & 0 & 0 & 0   & 0 \\
%   0 & 1 & 0 & 0 & .8 & 0 \\
%   0 & 0 & 0 & 0 & 1   & 0 \\
%   0 & 0 & 0 & 0 & 0   & 1  
% \end{array} \right)
% \end{equation} 

The \PageRank \ scores assigned to nodes in that digraph are: 
$$
\mathcal{P}(1) = \mathcal{P}(3) = \mathcal{P}(4) = 0.0958,\;\; \mathcal{P}(2) = 0.1410,\;\; 
\mathcal{P}(5) = 0.2133,  
$$
$$
{\rm and} \;\; \mathcal{P}(6) = 0.3585. 
$$

The individuals \lq 2\rq \ and \lq 5\rq \ are now untied, and we have better grounds to trust Programmer \lq 5\rq \ over Programer \lq 2\rq. 

%--------------------
%--------------------

\begin{figure}[htbp]
\begin{center}
%	 \centering
 	\includegraphics[width=0.9\textwidth]{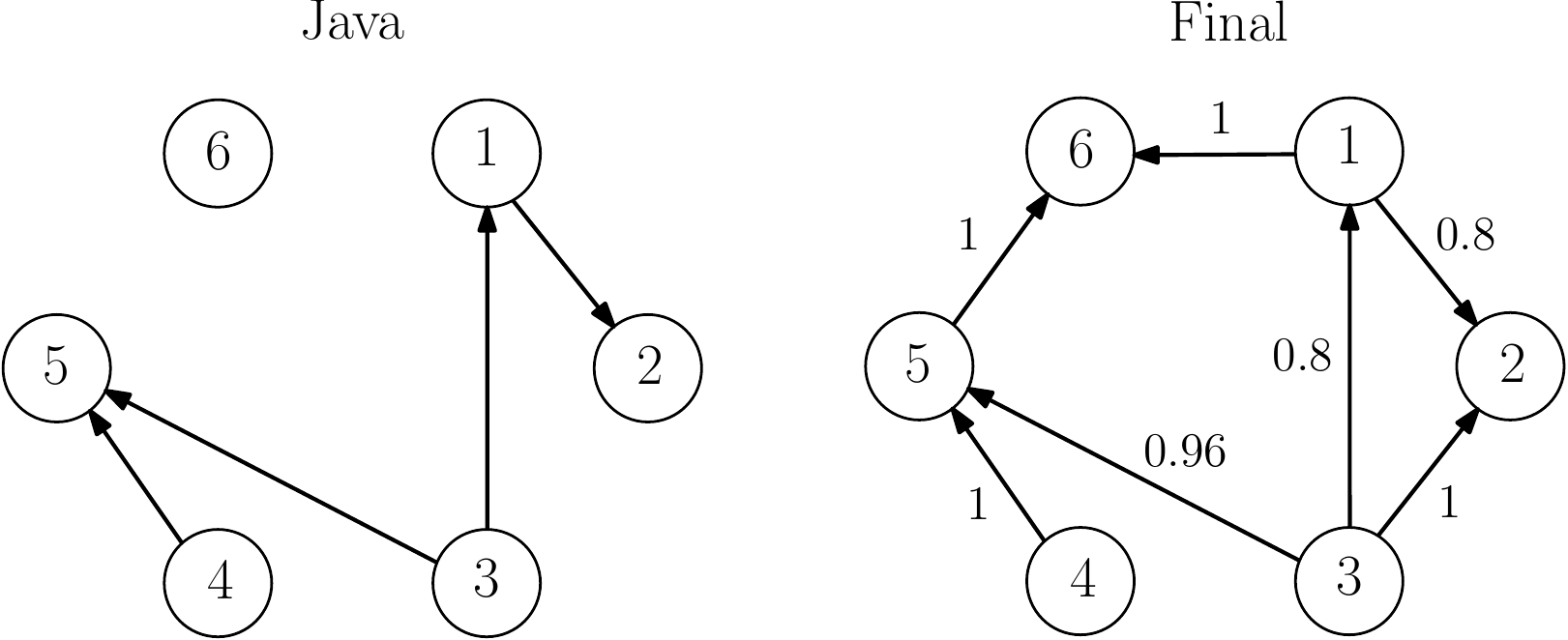}
	 \caption{Endorsements for \lq Java\rq \ (left), and endorsements for \lq Programming\rq, with information deduced from \lq C++\rq \ and \lq Java\rq \ (right).}
	 \label{fig:endorsements3}
\end{center}	
\end{figure}

Let us now suppose that the endorsement digraph for \lq Java\rq \ is the one given in Figure~\ref{fig:endorsements3} (left). We can include the endorsements for \lq Java\rq in the same manner: %or any other skill that is related to \lq Programming\rq. 

$$
\mathbf{Q}_2 = \mathbf{Q}_{1} + \pi_{2,0} ((\mathbf{J}^{(6)} - \mathbf{Q}_{1}) \circ \mathbf{M}_2), 
$$

where again $\pi_{2,0} = 0.8$. The result is given in Figure~\ref{fig:endorsements3} (right). 

%--------------------
%-------------------- 

If we apply \PageRank \ to this final digraph we get: 
$$
\mathcal{P}(1) = 0.1178, \;\; \mathcal{P}(2) = 0.1681,\;\; \mathcal{P}(3) = \mathcal{P}(4) = 0.0945,\;\;   
$$
$$
\mathcal{P}(5) = 0.2027, \;\; {\rm and} \;\; \mathcal{P}(6) = 0.3224. 
$$

% and we get
% \begin{equation}
% \label{eq:q2}
% \mathbf{Q}_2 = \left( \begin{array}{cccccc}
% 0 & 0 & 0 & 0 & 0   & 1 \\
% 0 & 0 & 0 & 0 & 0   & 0 \\
% 0 & 1 & 0 & 0 & .8 & 0 \\
% 0 & 0 & 0 & 0 & 1   & 0 \\
% 0 & 0 & 0 & 0 & 0   & 1  
% \end{array} \right)
% \end{equation} 

With the aid of the new endorsements, Programmer \lq 1\rq \ differentiates itself from Programmers \lq 3\rq \ and \lq 4\rq. 

%--------------------------
\subsection{Some properties}
%--------------------------

We now pay attention to some properties of endorsement deduction:

\begin{proposition}
\label{prop:properties}
Let the matrices $\mathbf{Q}_k$ be defined as in Equations \ref{eq:deductive}. Then, for all $1 \leq i,j \leq n$, the entry $\mathbf{Q}_k(i,j)$ satisfies the following properties:
\begin{enumerate}[a)]
\item $0 \leq \mathbf{Q}_k(i,j) \leq 1$, for all $0 \leq k \leq \ell$.
\item $\mathbf{Q}_k(i,j) \geq \mathbf{Q}_{k-1}(i,j)$, for all $1 \leq k \leq \ell$.
\item $\mathbf{Q}_\ell(i,j) = 0$ if, and only if, $\mathbf{M}_k(i,j) = 0$ for all $0 \leq k \leq \ell$. 
\item $\mathbf{Q}_\ell(i,j) = 1$ if, and only if, there exists some skill $k$, with $0 \leq k \leq \ell$, such that $\mathbf{M}_k(i,j) = 1$, and $\pi_{k,0} = 1$. In particular, if $\mathbf{M}_0(i,j) = 1$, then $\mathbf{Q}_\ell(i,j) = 1$, since $\pi_{0,0} = 1$.
\end{enumerate}
\end{proposition}

We omit the proofs, as they follow from the straightforward application of Equations \ref{eq:deductive}, and previous definitions. 

Put into plain words, Proposition \ref{prop:properties} states that a particular entry $\mathbf{Q}_k(i,j)$ can only grow with $k$, up to a limit of $1$. This maximum can only be reached if $i$ endorses $j$ directly for skill $0$, or for some other skill that implies skill $0$ with probability $1$. No other endorsement can have the same effect. 

This implies that, if two members of the network, $i$ and $j$, who were tied up in plain PageRank, get untied after deduction, it is because one of them has received additional endorsements for other skills that are related to the main skill (or has received more relevant endorsements). % Furthermore, the rank gained by a user after adding deduced endorsements with weight lower than $1$ will be less than 

%%======================================%%
%%             COMPARISON               %%
%%======================================%%
 
\section{Simple ranking vs. ranking with deduction}
\label{sec:comparison}

%--------------------------------
\subsection{Evaluation criteria}
\label{sec:criteria}
%--------------------------------

There is an extensive literature on the evaluation of information retrieval and ranking systems (see \cite{WIR13}, Sec. 1.2; \cite{Rob00}, and others). Several evaluation criteria and measures have been developed for that purpose, such as \emph{precision}, \emph{recall}, \emph{$F$-measure}, \emph{average precision}, \emph{P@n}, etc. All these measures rely on a set of assumptions, which include, among others, the existence of: 

\begin{enumerate}
\item a benchmark collection $E$ of personal profiles (potential experts), 
\item a benchmark collection $S$ of skills,
\item a (total binary) judgement function $r: E \times S \rightarrow \{ 0,1 \}$, stating whether a person $e \in E$ is an expert with respect to a skill $s \in S$. 
\end{enumerate}

The above conditions have been taken from \cite{WIR13}, Sec. 1.2, and adapted to our situation. Unfortunately, none of these assumptions applies in our case. 

To the best of our knowledge, there does not exist any reliable open-access ground-truth dataset of experts and skills, connected by endorsement relations. To begin with, the endorsement feature is relatively new, and still confined to a few social networks (e.g. \LinkedIn \ and \ResearchGate). In \ResearchGate \ in particular, it was only introduced in February, 2013, and not enough data has accumulated so far. On the other hand, \LinkedIn \ does not disclose sensitive information of its members (including their contacts or their endorsements), due to privacy concerns. Crawling the network, such as in \cite{crowd} is not allowed: \LinkedIn's terms of use specifically prohibit to \emph{\lq\lq scrape or copy profiles and information of others through any means (including crawlers, browser plugins and add-ons, and any other technology or manual work)\rq\rq} (see \cite{user-agreement}). Therefore, assumptions 1 and 2 do not hold in our case.  

The third assumption is also problematic: Even if we had a dataset with endorsements, we would still need a \lq higher authority\rq, or an \lq oracle\rq, to judge about the expertise of a person. Moreover, since our goal is to rank experts, a binary oracle would not suffice. 

Traditionally, ranking methods have been validated by carrying out surveys among a group of users (e.g. \cite{blogranker06}), which in our opinion, is very subjective and error-prone. We propose a more objective validation methodology, which is based on the following criteria: 

\begin{enumerate}
\item \emph{Sanity check:} Our ranking with deduction is close to the ranking provided by \PageRank. If we use endorsement deduction in connection with \PageRank, results should not differ too much from \PageRank; i.e. our method should only modulate the ranking provided by \PageRank \ by introducing local changes. In order to evaluate the closeness between two rankings we can use some measure of rank correlation. Measures of rank correlation have been studied for more than a century now, and the best known of them are Spearman's correlation coefficient $\rho$ \cite{Spe04}, and Kendall's correlation coefficient $\tau$ \cite{Ken38}. 
\item Ranking with deduction results in less ties than \PageRank. Ties are an expression of ambiguity, hence a smaller number of ties is desirable. In the example of Section \ref{sec:deductive}, we have seen that ranking with deduction resolves a tie produced by \PageRank. However, this has to be confirmed by meaningful experiments. 
\item Ranking with deduction is more robust than \PageRank \ to \emph{collusion spamming}. Collusion spamming is a form of \emph{link spamming}, i.e. an attack to the reputation system, whereby a group of users collude to create artificial links among themselves, and thus manipulate the results of the ranking algorithm, with the purpose of getting higher reputation scores than they deserve  \cite{alliances05,taxonomy05}. If the users create false identities (or duplicate identities) to carry out the spamming, the strategy is known as \emph{Sybil attack} \cite{Dou02}. 
\end{enumerate}

%-----------------------------------------
\subsection{Experimental setup and results}
\label{sec:setup}
%-----------------------------------------

We now proceed to evaluate our ranking with deduction, according to each of the above criteria. Our experimental benchmark consists of a randomly generated social network that replicates some features of \LinkedIn \ at a small scale \cite{synthetic}. \LinkedIn \ consists of an undirected  \emph{base network} $\mathcal(L)$, or \emph{network of contacts}, and for each skill, the corresponding endorsements form a directed subgraph of $\mathcal(L)$. In \cite{Les08}, Leskovec formulates a model that describes the evolution of several social networks quite accurately, including \LinkedIn, although this model is limited to the network of contacts $\mathcal(L)$, and does not account for the endorsements, since that feature was introduced in \LinkedIn \ later. 
% This is due to the way that LinkedIn operates: In order to establish a contact with another user, he/she must accept the contact request, thus creating a bidirectional link. On the other hand, one can endorse another user without expecting 

\begin{figure}[htbp]
\begin{center}
%	 \centering
 	\includegraphics[width=1.0\textwidth]{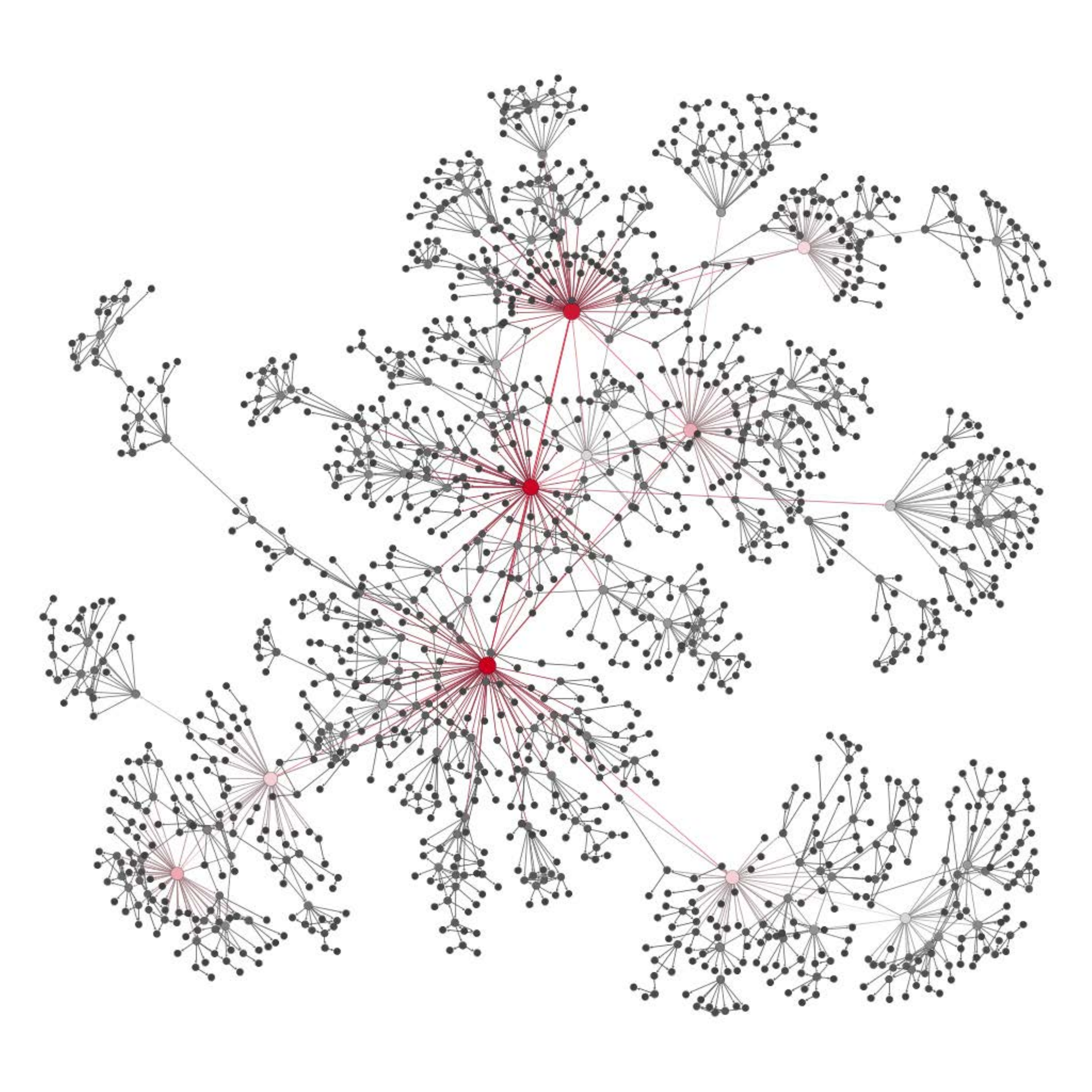}
	 \caption{Base network of 1493 nodes, used for experiments}
	 \label{fig:miniLinkedIn}
\end{center}	
\end{figure}

Thus, we have implemented Leskovec's model and used it to generate an undirected network of contacts with $1 493$ nodes and $2 489$ edges, represented in Figure \ref{fig:miniLinkedIn}. 

%--------------------
%--------------------

Additionally, we have considered five skills: 1. Programming, 2. C++, 3. Java, 4. Mathematical Modelling, 5. Statistics. We have chosen these skills for two main reasons: 
\begin{enumerate}
\item These five skills abound in a small \LinkedIn \ community consisting of 278 members, taken  from our \LinkedIn \ contacts, which we have used as a sample to collect some statistics. 
\item These five skills can be clearly separated into two groups or clusters, namely programming-related skills, and mathematical skills, with a large intra-cluster correlation, and a smaller inter-cluster correlation. This is a small-scale representation of the real network, where skills can be grouped into clusters of related skills, which may give rise to different patterns of interaction among skills. 
\end{enumerate}

We have computed the co-occurrences of the five skills in our small community, resulting in the co-occurrence matrix $\mathbf{\Pi}_1$ of Eq. \ref{eq:obvserved1}. The entry $\mathbf{\Pi}_1 (i, j)$ is the ratio between the number of nodes that have been endorsed for both skills, $i$ and $j$, and the number of nodes that have been endorsed for skill $i$ alone. %It may be taken as an estimate of the conditional probability 

\begin{equation}
\label{eq:obvserved1}
\mathbf{\Pi}_1 = \left( \begin{array}{ccccc}
1 & 0.42 & 0.42 & 0.5 & 0.33 \\
0.62 & 1 & 0.62 & 0.25 & 0.12 \\
0.62 & 0.62 & 1 & 0.12 & 0.12 \\
0.75 & 0.25 & 0.12 & 1 & 0.5 \\
0.5 & 0.12 & 0.12 & 0.5 & 1 
\end{array} \right)
\end{equation} 

Now, for each skill we have constructed a random endorsement digraph (a random sub-digraph of the base network), in such a way that the above co-occurrences are respected. We have also taken care to respect the relative endorsement frequency for each individual skill. 
The problem of constructing random endorsement digraphs according to a given co-occurrence matrix is not trivial, and may bear some  interest in itself \cite{synthetic}. We have chosen to skip the details here because it is not our main concern in the present paper. The base network and the endorsement digraphs can be downloaded from \url{http://www.cig.udl.cat/sitemedia/files/MiniLinkedIn.zip}.

Next, for each skill we have computed two rankings, one using the simple \PageRank \ algorithm, and another one using \PageRank \ with deduction. For \PageRank \ with deduction we have used the skill deduction matrix $\mathbf{\Pi}_2$ given in Eq. \ref{eq:dedumatrix1}. This matrix has been constructed by surveying a group of seven experts in the different areas involved. 

\begin{equation}
\label{eq:dedumatrix1}
\mathbf{\Pi}_2 = \left( \begin{array}{ccccc}
1 & 0.7 & 0.7 & 0.4 & 0.3 \\
1 & 1 & 0.6 & 0.4 & 0.3 \\
1 & 0.7 & 1 & 0.4 & 0.3 \\
0.3 & 0.2 & 0.2 & 1 & 0.8 \\
0.3 & 0.2 & 0.2 & 1 & 1 
\end{array} \right)
\end{equation}

\begin{figure}[htbp]
\begin{center}
%	 \centering
 	\includegraphics[width=0.5\textwidth]{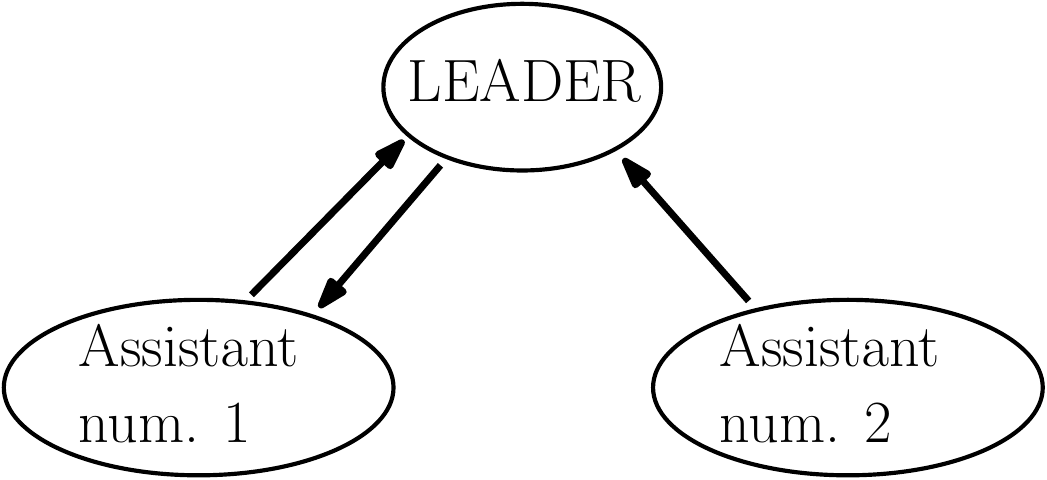}
	 \caption{Link spam alliance: Three people collude to promote one of them.}
	 \label{fig:alliance1}
\end{center}	
\end{figure}

For each skill we have computed the correlation between both rankings, and the number of ties in each case, according to the first two criteria described above. Additionally, in order to test the robustness of the method to collusion spamming, we have added to each endorsement digraph, a small community of new members (the \emph{cheaters}), who try to subvert the system by promoting one of them (their \emph{leader}) as an expert in the corresponding skill. We have chosen the most effective configuration for such a spamming community, as described in \cite{alliances05}, and depicted in Figure \ref{fig:alliance1}. Thereupon we have compared the position of the leader of cheaters in simple \PageRank \ with its position in \PageRank \ with deduction. 

%--------------------
%--------------------

%--------------------
\begin{table}[htp]
\footnotesize
%\tabcolsep=0.11cm  % The purpose of this is to control the width of the columns, so that it does not exceed margins
%\begin{center}
\scalebox{0.9}{ 
\tabcolsep=0.11cm
\begin{tabular}{|cc|*{2}{c}|*{2}{c}|*{3}{c}|*{3}{c}|} \hline
\multicolumn{2}{|c|}{} & \multicolumn{2}{c|}{\textbf{Number}} & \multicolumn{2}{c|}{\textbf{Correlat. }} & \multicolumn{3}{c|}{\textbf{Number of ties}} &  \multicolumn{3}{c|}{\textbf{Position of leader}}  \\

\multicolumn{2}{|c|}{} &  \multicolumn{2}{c|}{\textbf{of endor-}}  &  \multicolumn{2}{c|}{}  &  &  &  &  &  &  \\

\multicolumn{2}{|c|}{\textbf{Skill}} & \multicolumn{2}{|c|}{\textbf{sements}} & \textbf{$\rho$} & \textbf{$\tau$} & \textbf{without} & \textbf{with} & \textbf{\%} & \textbf{without} &  \textbf{with} & \textbf{\%} \\

\multicolumn{2}{|c|}{} &  \multicolumn{2}{|c|}{\textbf{(arcs)}}  &  &  & \textbf{ deduct. } & \textbf{ deduc. } & \textbf{ reduc.} & \textbf{ deduct. } & \textbf{ deduc. } & \textbf{ fall} \\ 
\hline  
\hline 
\multicolumn{2}{|c|}{Program. } & \multicolumn{2}{|c|}{220} & 0.89 & 0.76 & 1460 & 1316 & 10\% & 1 & 48 & 3\% \\ 
\multicolumn{2}{|c|}{C++} & \multicolumn{2}{|c|}{140} & 0.85 & 0.63 & 1478 & 1304 & 12\% & 4 & 48 & 3\% \\ 
\multicolumn{2}{|c|}{Java} & \multicolumn{2}{|c|}{137} & 0.85 & 0.63 & 1486 & 1292 & 13\% & 1 & 48 & 3\% \\
\multicolumn{2}{|c|}{Math Mod } & \multicolumn{2}{|c|}{134} & 0.85 & 0.63 & 1483 & 1318 & 11\% & 1 & 45 & 3\% \\
\multicolumn{2}{|c|}{Statistics} & \multicolumn{2}{|c|}{128} & 0.85 & 0.63 & 1486 & 1304 & 12\% & 1 & 45 & 3\% \\
\hline
\multicolumn{2}{|c|}{\textbf{AVG}} & \multicolumn{2}{|c|}{} &  &  &  &  & \textbf{11.6\%} &  &  & \textbf{3\%} \\
% \multicolumn{2}{|c|}{\textbf{STD}} & \textbf{571.2} & \textbf{292.7} & \textbf{2 715.5} & \textbf{2 600} & \textbf{3.36} & \textbf{0.74} \\
\hline
\end{tabular} }
\vspace{5mm}
\caption{Results from the first experiment}
\label{tab:results1}
%\end{center}
\end{table}
%--------------------

Table \ref{tab:results1} summarizes the results of the aforementioned experiments. Regarding the first criterion, we can see that there is a very high correlation between \PageRank \ with deduction and \PageRank \ without deduction for all skills. According to Kendall's $\tau$, there is a significant agreement between both rankings, with a significance level of $0.001$, or even better. Spearman's $\rho$ shows an even higher agreement. 

%The smallest value obtained corresponds to Kendall's $\tau$ in the last four skills, probably due to the smaller number of arcs in the corresponding endorsement digraphs. For the sake of rigour, we may perform a significance test for this particular value, in order to confirm whether we can reject the null hypothesis. Nevertheless, critical values of ranking correlation coefficients are usually tabulated up to a small number of elements $n$, typically $30$ or $50$, which is very far from our $1493$ nodes. Take for instance the table in \cite{tables1}: We can see that for $60$ elements (the maximum in the table), the upper critical value of Kendall's $\tau$ for a significance level of $0.001$ is $0.272$. Considering that critical values decrease as $n$ increases, and that our value of $0.63$ exceeds $0.272$ by $0.358$, we can safely reject the null hypothesis and conclude that there is a significant agreement between both rankings, with a significance level of $0.001$, or even better.

\begin{figure}[htbp]
\begin{center}
%	 \centering
 	\includegraphics[width=0.95\textwidth]{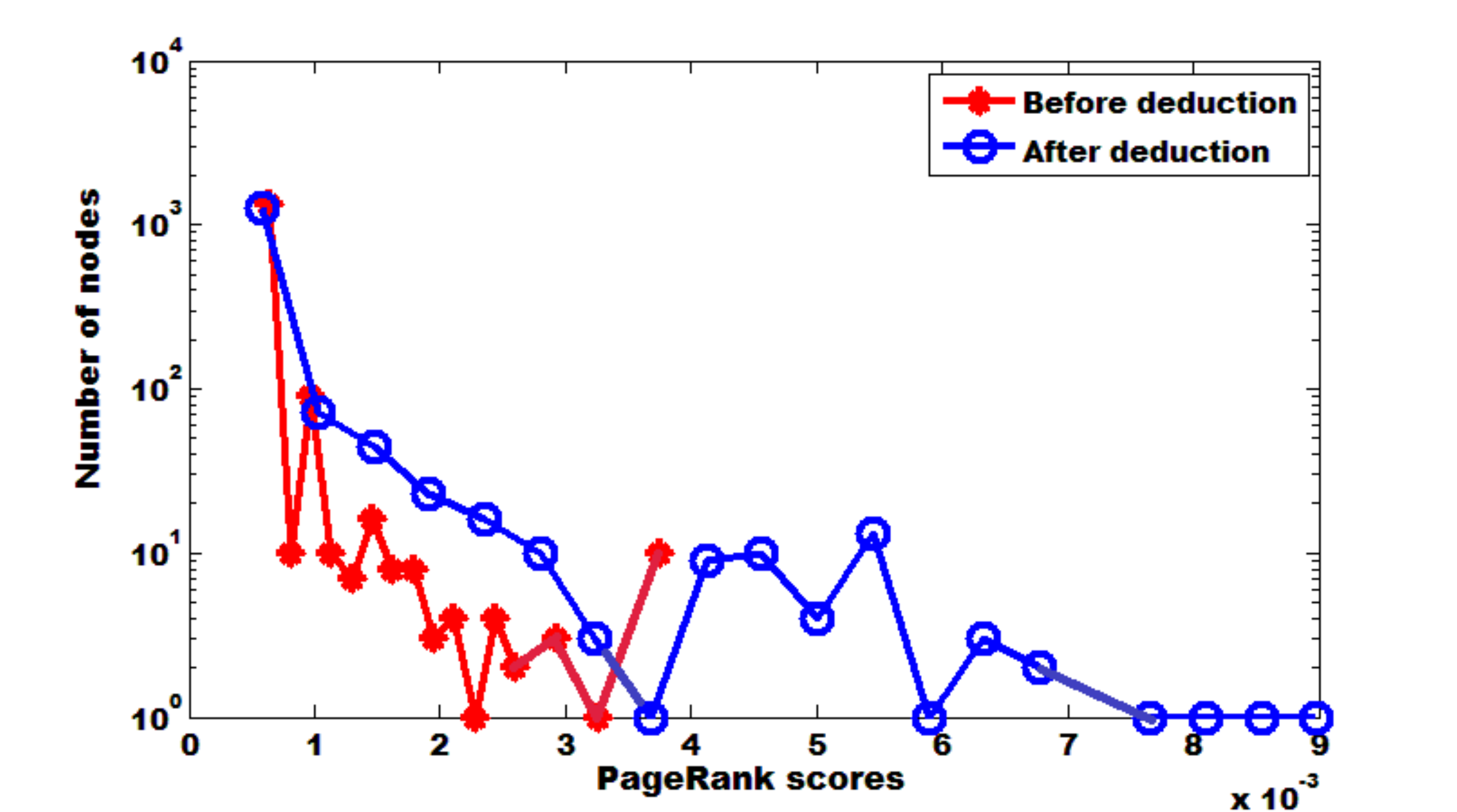}
	 \caption{Histograms of \PageRank \ scores, before and after deduction}
	 \label{fig:histograms}
\end{center}	
\end{figure}

With respect to the second criterion, the experiments also yield unquestionable results: For all skills, the number of ties is significantly reduced. This is also reflected in the distribution of \PageRank \ scores, shown in Fig. \ref{fig:histograms}. After deduction, the scores are more evenly distributed. 

%--------------------
%--------------------

As for the third criterion, in all cases there is a detectable drop in the position of the leader of cheaters, which may lead us to conclude that \PageRank \ with deduction is more robust to collusion spam than simple \PageRank. However, this may not lead us to the conclusion that \PageRank \ with deduction is an effective mechanism against collusion spam. Actually, the spam alliance that we have introduced in our experiments is rather weak. If we strengthen the spam alliance, then \PageRank \ with deduction may also be eventually deceived. Table \ref{tab:alliances} illustrates the effect of strengthening the spam alliance, by increasing the number of assistants from 2 to 8. For each spam alliance there are three columns, labelled as \lq -\rq \ (position of the leader in the ranking without deduction), \lq +\rq \ (position in the ranking with deduction), and \lq \%\rq \ (effectiveness of deduction, measured as the leader's fall in position, in percentage points). 

%--------------------
\begin{table}[htp]
\footnotesize
%\tabcolsep=0.11cm
%\begin{center}
%\scalebox{0.9}{
\tabcolsep=0.11cm
\begin{tabular}{|cc|*{3}{c}|*{3}{c}|*{3}{c}|*{3}{c}|*{3}{c}|*{3}{c}|*{3}{c}|} \hline
\multicolumn{2}{|c|}{} & \multicolumn{21}{c|}{\textbf{Number of assistants}} \\
\multicolumn{2}{|c|}{} & \multicolumn{21}{c|}{} \\ % -- Blank line

\multicolumn{2}{|c|}{} & \multicolumn{3}{c|}{\textbf{2}} & \multicolumn{3}{c|}{\textbf{3}} & \multicolumn{3}{c|}{\textbf{4}} & \multicolumn{3}{c|}{\textbf{5}} & \multicolumn{3}{c|}{\textbf{6}} & \multicolumn{3}{c|}{\textbf{7}} & \multicolumn{3}{c|}{\textbf{8}} \\

\multicolumn{2}{|c|}{\textbf{Skill}} &  \multicolumn{3}{c|}{}  &  \multicolumn{3}{c|}{}  & \multicolumn{3}{c|}{} & \multicolumn{3}{c|}{} & \multicolumn{3}{c|}{} & \multicolumn{3}{c|}{} & \multicolumn{3}{c|}{} \\

\multicolumn{2}{|c|}{} & \textbf{-} & \textbf{+} & \textbf{\%} & \textbf{-} &  \textbf{+} & \textbf{\%} & \textbf{-} &  \textbf{+} & \textbf{\%} & \textbf{-} & \textbf{+} & \textbf{\%} & \textbf{-} &  \textbf{+} & \textbf{\%} & \textbf{-} &  \textbf{+} & \textbf{\%} & \textbf{-} &  \textbf{+} & \textbf{\%}\\

%\multicolumn{2}{|c|}{} &  \multicolumn{2}{|c|}{\textbf{(arcs)}}  &  &  & \textbf{ deduct. } & \textbf{ deduc. } & \textbf{ reduc.} & \textbf{ deduct. } & \textbf{ deduc. } & \textbf{ fall} \\ 
\hline  
\hline 
\multicolumn{2}{|c|}{Program. } & 1 & 48 & 3 & 1 & 31 & 2 & 1 & 10 & 1 & 1 & 6 & 0 & 1 & 4 & 0 & 1 & 2 & 0 & 1 & 1 & 0\\ 
\multicolumn{2}{|c|}{C++} & 4 & 48 & 3 & 3 & 29 & 2 & 3 & 11 & 1 & 1 & 6 & 0 & 1 & 5 & 0 & 1 & 2 & 0 & 1 & 1 & 0\\ 
\multicolumn{2}{|c|}{Java} & 1 & 48 & 3 & 1 & 30 & 2 & 1 & 10 & 1 & 1 & 6 & 0 & 1 & 4 & 0 & 1 & 2 & 0 & 1 & 1 & 0\\
\multicolumn{2}{|c|}{Math Mod } & 1 & 45 & 3 & 1 & 27 & 2 & 1 & 12 & 1 & 1 & 4 & 0 & 1 & 2 & 0 & 1 & 2 & 0 & 1 & 1 & 0\\
\multicolumn{2}{|c|}{Statistics} & 1 & 45 & 3 & 1 & 28 & 2 & 1 & 11 & 1 & 1 & 4 & 0 & 1 & 2 & 0 & 1 & 2 & 0 & 1 & 1 & 0\\
% \hline
% \multicolumn{2}{|c|}{\textbf{AVG}} &  &  &  &  &  & \textbf{47\%} &  &  & \textbf{8\%} \\
% \multicolumn{2}{|c|}{\textbf{STD}} & \textbf{571.2} & \textbf{292.7} & \textbf{2 715.5} & \textbf{2 600} & \textbf{3.36} & \textbf{0.74} \\
\hline
\end{tabular} %}
\vspace{5mm}
\caption{Effect of endorsement deduction in the presence of different spam alliances}
\label{tab:alliances}
%\end{center}
\end{table}
%--------------------

The simplest way to implement a collusion spam attack is the so-called Sybil attack, in which a single attacker creates several fake identities, without the need to collaborate with other people. Proactive measures against the Sybil attack focus on limiting the capability of a malicious user to create a large amount of accounts. It has been proven that a trusted central authority issuing credentials unique to an actual human being is the only method that may eliminate Sybil attacks completely \cite{Dou02}. Requiring fees per identity could mitigate them when the cost of the accounts is larger than the benefit for the attacker. Reactive measures try to mitigate the effect of such an attack. The SybilGuard \cite{SybilGuard06} proposal bounds the number of identities a malicious user can create under the assumption that a malicious user could create many identities but few trust relations, so that there exists a small cut in the graph between Sybil nodes and honest ones.

Our proposal belongs to the second category. It is not designed to be a safeguard against the Sybil attack, but the experiments have shown that it provides some reactive measure behaviour. In any case, the social network platform implementing our proposal could include all the proactive and reactive measures against sybil attacks without interferring with our method. A complete survey of attack and defense techniques for reputation  systems is given in \cite{reput}. 
% Several effective mechanisms have been proposed to fight collusion spam, an example being the so-called \emph{asymmetric reputation systems}. A complete survey of such systems is given in \cite{reput}. Presumably, these mechanisms will give better results when combined with deduction.

On the other hand, our endorsement digraphs are rather sparse, since our contacts are rather lazy when it comes to endorsing each other.  It is reasonable to predict that if we should consider more skills, and if the total number of endorsements should increase, then the effects of \PageRank \ with deduction will be stronger. 

In order to verify this prediction, we have carried out a second experiment on the same base network. For practical reasons we have decided to keep the set of skills invariant for the moment, and increase the number of endorsements. Thus we have generated a second set of endorsement digraphs, with a larger number of arcs. This time we cannot enforce the co-occurrences observed in our small \LinkedIn \ community. The co-occurrence matrix obtained is given in Eq. \ref{eq:exper2} for the sake of comprehensiveness.

\begin{equation}
\label{eq:exper2}
\mathbf{\Pi}_3 = \left( \begin{array}{ccccc}
 1   & 0.88 & 0.87 &  1   & 0.61 \\
0.32 &  1   & 0.9  & 0.73 & 0.61 \\
0.31 & 0.89 &  1   & 0.63 & 0.59 \\
0.42 & 0.85 & 0.75 &  1   & 0.7  \\
0.24 & 0.67 & 0.66 & 0.66 & 1 
\end{array} \right)
\end{equation}

Subsequently we have performed the same computations on this second set of endorsement digraphs, obtaining the results recorded in Table \ref{tab:results2}. These results fully confirm our prediction: There is an increase in the correlation coefficients (except in one case), as well as a larger reduction in the number of ties, and a more significant fall in the position of the leader of cheaters. 

%--------------------
\begin{table}[htp]
\footnotesize
%\tabcolsep=0.11cm
%\begin{center}
\scalebox{0.9}{
\tabcolsep=0.11cm
\begin{tabular}{|cc|*{2}{c}|*{2}{c}|*{3}{c}|*{3}{c}|} \hline
\multicolumn{2}{|c|}{} & \multicolumn{2}{c|}{\textbf{Number}} & \multicolumn{2}{c|}{\textbf{Correlat. }} & \multicolumn{3}{c|}{\textbf{Number of ties}} &  \multicolumn{3}{c|}{\textbf{Position of leader}}  \\

\multicolumn{2}{|c|}{} &  \multicolumn{2}{c|}{\textbf{of endor-}}  &  \multicolumn{2}{c|}{}  &  &  &  &  &  &  \\

\multicolumn{2}{|c|}{\textbf{Skill}} & \multicolumn{2}{|c|}{\textbf{sements}} & \textbf{$\rho$} & \textbf{$\tau$} & \textbf{without} & \textbf{with} & \textbf{\%} & \textbf{without} &  \textbf{with} & \textbf{\%} \\

\multicolumn{2}{|c|}{} &  \multicolumn{2}{|c|}{\textbf{(arcs)}}  &  &  & \textbf{ deduct. } & \textbf{ deduc. } & \textbf{ reduc.} & \textbf{ deduct. } & \textbf{ deduc. } & \textbf{ fall} \\ 
\hline  
\hline 
\multicolumn{2}{|c|}{Program. } & \multicolumn{2}{|c|}{427} & 0.76 & 0.63 & 1428 & 625 & 56\% & 1 & 175 & 12\% \\ 
\multicolumn{2}{|c|}{C++} & \multicolumn{2}{|c|}{1793} & 0.97 & 0.93 & 1005 & 575 & 43\% & 66 & 178 & 7\% \\ 
\multicolumn{2}{|c|}{Java} & \multicolumn{2}{|c|}{1856} & 0.97 & 0.93 & 1005 & 566 & 44\% & 63 & 180 & 8\% \\
\multicolumn{2}{|c|}{Math Mod } & \multicolumn{2}{|c|}{1406} & 0.95 & 0.89 & 1130 & 652 & 42\% & 56 & 168 & 7\% \\
\multicolumn{2}{|c|}{Statistics} & \multicolumn{2}{|c|}{1447} & 0.96 & 0.90 & 1113 & 580 & 48\% & 58 & 169 & 7\% \\
\hline
\multicolumn{2}{|c|}{\textbf{AVG}} & \multicolumn{2}{|c|}{} &  &  &  &  & \textbf{47\%} &  &  & \textbf{8\%} \\
% \multicolumn{2}{|c|}{\textbf{STD}} & \textbf{571.2} & \textbf{292.7} & \textbf{2 715.5} & \textbf{2 600} & \textbf{3.36} & \textbf{0.74} \\
\hline
\end{tabular} }
\vspace{5mm}
\caption{Results from the second experiment}
\label{tab:results2}
%\end{center}
\end{table}
%--------------------

%-----------------------------------------
% \subsection{Analysis of results}
% \label{sec:analysis}
%-----------------------------------------

%%======================================%%
%%                 OPEN                 %%
%%======================================%%
 
\section{Conclusions and future research}
\label{sec:open}

In this paper we describe endorsement deduction, a pre-processing algorithm to enrich the endorsement digraphs of a social network with endorsements, such as \LinkedIn \ or \ResearchGate, which can then be used in connection with a ranking method, such as \PageRank, to compute an authority score of network members with respect to some desired skill. Endorsement deduction makes use of the relationships among the different skills, given by an ontology in the form of a \emph{skill deduction matrix}. A preliminary set of experiments shows that the rankings obtained by this method do not differ much from the rankings obtained by simple \PageRank, and that this method represents an improvement over simple \PageRank \ with respect to two additional criteria: number of ties, and robustness to collusion spam. 

%extend the \PageRank \ method to compute authority scores among users of a social network with endorsements, such as \LinkedIn \ or \ResearchGate, with respect to some desired skill. Our extension makes use of the relationships among the different skills, given by a \emph{skill deduction matrix}. 

Our experiments also show that the benefits provided by \PageRank \ with deduction are likely to increase in the future, with the densification of the endorsement networks, and the introduction of new skills. However, this has to be confirmed by larger-scale experiments. It could also be interesting to test our method with other ranking algorithms, such as \Hits.  

Although \LinkedIn \ and \ResearchGate \ are perhaps the best examples at hand, this system can also be extended to other social networks and platforms. Take, for instance, the open access publishing platform arXiv\footnote{\url{http://arxiv.org}}. In order to submit a paper on a particular topic, say \lq Statistics\rq, an author has to be endorsed by another trusted author for \lq Statistics\rq. However, it might as well be possible to allow an author submit a paper on \lq Statistics\rq \ if he/she has been endorsed for \lq Probability Theory\rq.  

There are many Internet forums, such as the \lq StackExchange\rq \ suite, that assign an authority score to their members. As an illustration, let us pick one of the most popular forums of this family: The \MathStackExchange\footnote{\url{http://math.stackexchange.com}}, where users can pose questions and obtain answers about mathematical problems. As of today (July, 2015), the site has more than 152 000 members, and more than 467 000 questions have been posed. Members get credit points for the questions, answers, or comments that they post, via the votes of other members. A high authority score entitles a member to certain privileges. By design, all the votes are worth the same number of points, but a more realistic model would be to make the value of the votes dependent from the authority score of the voting person. Additionally, authority scores could be disassembled into areas of knowledge, since questions are tagged with the areas to which they belong (e.g. \lq Linear Algebra\rq, \lq Calculus\rq, \lq Probability\rq, etc.). 

\MathOverflow\footnote{\url{http://mathoverflow.net}} is very similar to \MathStackExchange, but it focuses on more technical questions in state-of-the-art mathematics. Due to its more \lq elitist\rq \ nature, \MathOverflow \ is smaller than the \MathStackExchange. Nevertheless, it is also a success story, with its more than 37 000 users, and \emph{circa} 62 000 questions posted, and it has become an undisputable actor in mathematical research, having attracted some of the world's top mathematicians \cite{klarreich11}. 

A competitor to \ResearchGate \ is \Academia, another academic social network designed to disseminate research results, ask and answer questions, and find like-minded collaborators. In both platforms, users can upload their papers, and tag them with the different research topics. Questions can be tagged too. It has been argued that, for the moment, these platforms have failed to attract some of the most experienced scientists. This may be partly due to the fact that scientists are conservative when it comes to adopting new technologies, but judging from \Egomnia's experience, it may also be partly due to the unreliability (or outright inexistence) of scoring and ranking mechanisms \cite{lugger12}. It would not be difficult for \ResearchGate \ to make the RG score more reliable by adopting the techniques discussed above. As a starting point, it would be interesting to extend the experimental analysis of Section \ref{sec:comparison} to a \ResearchGate-like network, and compare the results with the ones obtained here. 

It is worth observing that all these ideas are also applicable outside the academic realm. In principle, even \Google's search engine could make use of these techniques to find content by synonyms. \footnote{\Google \ already has some functionality for synonyms via the \lq $\sim$\rq \ operator.} In order to do that, they would need a very large semantic network, comprising all the potential keywords and their correlation. 

Similarly, every social network or video repository displays some featured content on the start page, whose popularity has been computed on the basis of the number of votes (i.e. clicks on the \lq Like\rq \ button). Yet, this content is usually tagged by topic, and hence, it might be possible to compute a more specific popularity score, and thus display content specifically tuned to the user's interests. The authors in \cite{weblogs06} propose a method for ranking weblogs. Their proposal consists in aggregating links by considering similarity in authors and topics between pairs of blogs. 

% In the past months, all the main social networks have been investing heavily in intelligent tools to expand their capabilities \cite{Smi14}, and a tool like this, which enhances the search capabilities of the network, might be most welcome. 

%Academia.edu\footnote{\url{http://http://www.academia.edu}: 3,224,803 academics have signed up to Academia.edu, adding 1,643,213 papers and 753,813 research interests. Academia.edu attracts over 5 million unique visitors a month.
%Youtube features the most popular videos, on the basis of the number of votes, or the number of views. Yet, videos are tagged by topic (e.g. folk music, basketball, etc.)

In any case, for the practical implementation of these techniques, two obvious problems arise: 

\begin{enumerate}
\item The first problem has to do with the estimation of the skill deduction matrix, which in this paper we estimate by polling a group of experts. There may be several alternatives to estimate the matrix from the social network itself, and it may be necessary to ponder the pros and cons of each alternative. %Section \ref{sec:correlation} suggests some alternatives for that purpose, but which one is more reliable in practice?
\item The second problem has to do with the feasibility of the computation. Assuming that the skill deduction matrix is available, computation of the weighted \PageRank \ is a costly process for a large social network. Methods for accelerating \PageRank \ computations have been considered in \cite{extrapolation,sparse}; we might need to adapt them to our situation. Additionally, we might need methods to accelerate the computation of the accumulated endorsement matrix $\mathbf{Q}$. 
\end{enumerate}

A subsidiary problem has to do with modelling the dynamics of endorsements in both \LinkedIn \ and \ResearchGate. 

%%======================================%%
%%======================================%%

\section*{Acknowledgements}

Authors have been partially funded by the Spanish Ministry of Economy and Competitiveness under projects TIN2010-18978,  IPT-2012-0603-430000, and MTM2013-46949-P. The first two authors wish to dedicate this paper to the memory of our friend and colleague Josep Maria Rib\'o Balust,who could not live to see the conclusion of his work.% We thank the anonymous referees for their timely and constructive comments, which have helped us improve the paper significantly. 

%%======================================%%
%%             REFERENCES               %%
%%======================================%%

%%==============================================%%
%%          	 LINKEDIN'S ENDORSEMENT           %%
%%==============================================%%
\section*{Appendix: The endorsement mechanism and its inconsistencies}

\begin{figure}[htbp]
\begin{center}
%	 \centering
 	\includegraphics[width=0.75\textwidth]{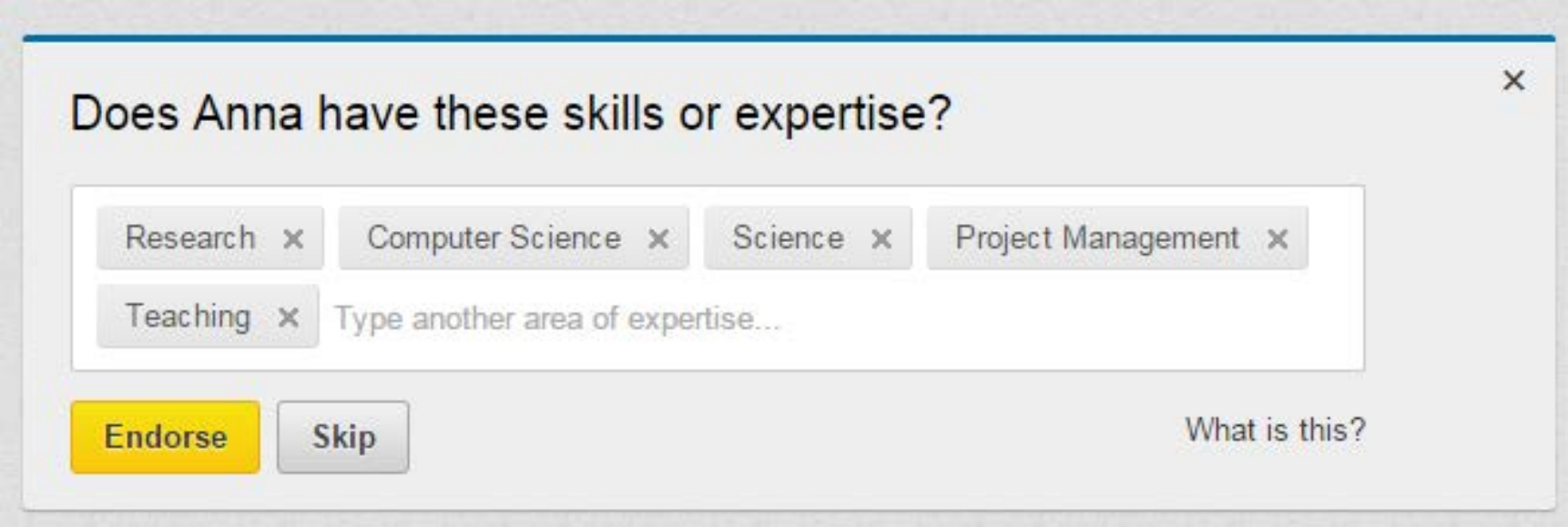}
	 \caption{First endorsement mechanism: Batch endorsement}
	 \label{fig:batch}
\end{center}	
\end{figure}

\begin{figure}[htbp]
\begin{center}
%	 \centering
 	\includegraphics[width=0.75\textwidth]{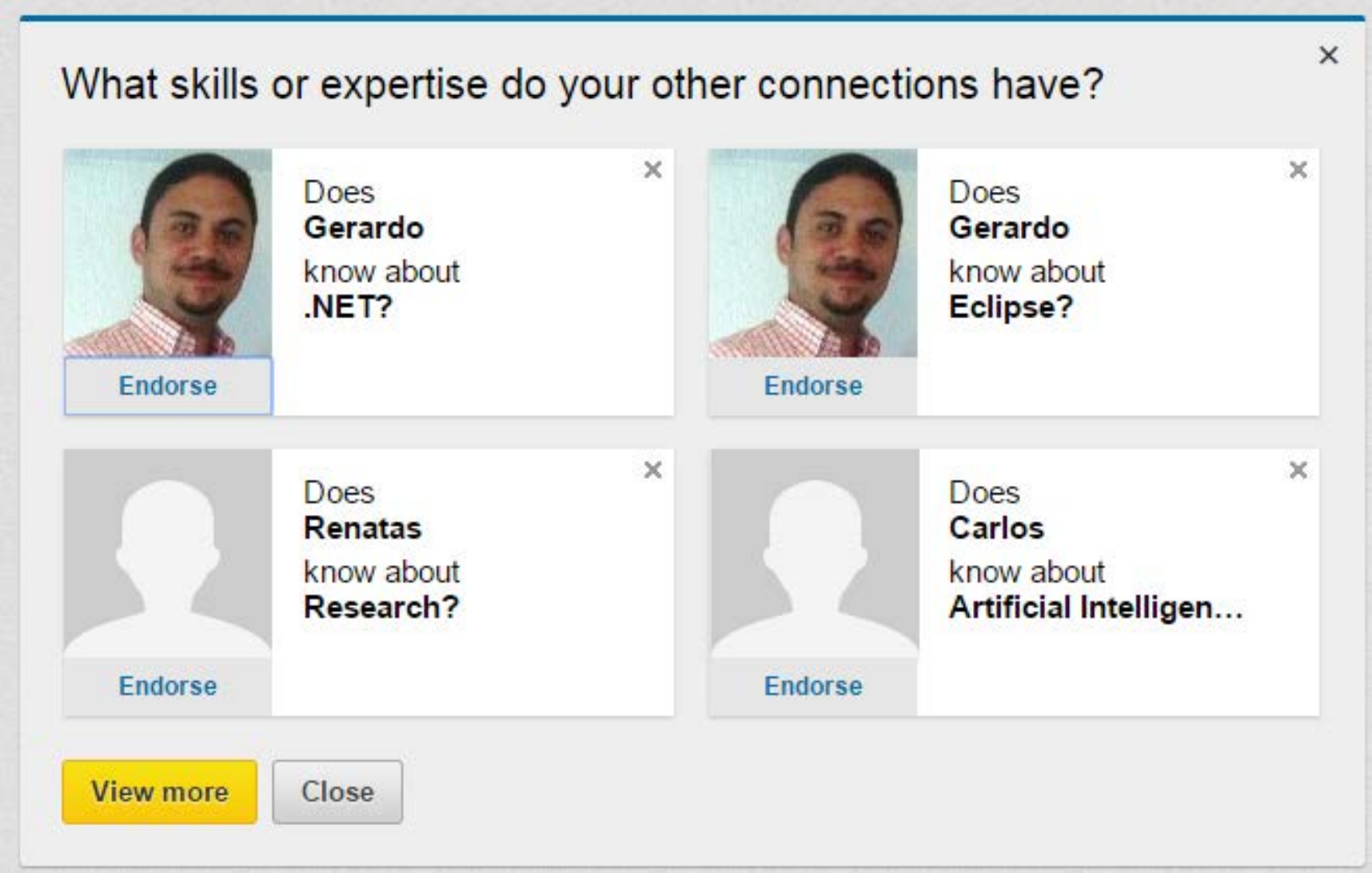}
	 \caption{Second endorsement mechanism: A group of four candidates to be endorsed}
	 \label{fig:four}
\end{center}	
\end{figure}

\begin{figure}[htbp]
\begin{center}
%	 \centering
 	\includegraphics[width=0.75\textwidth]{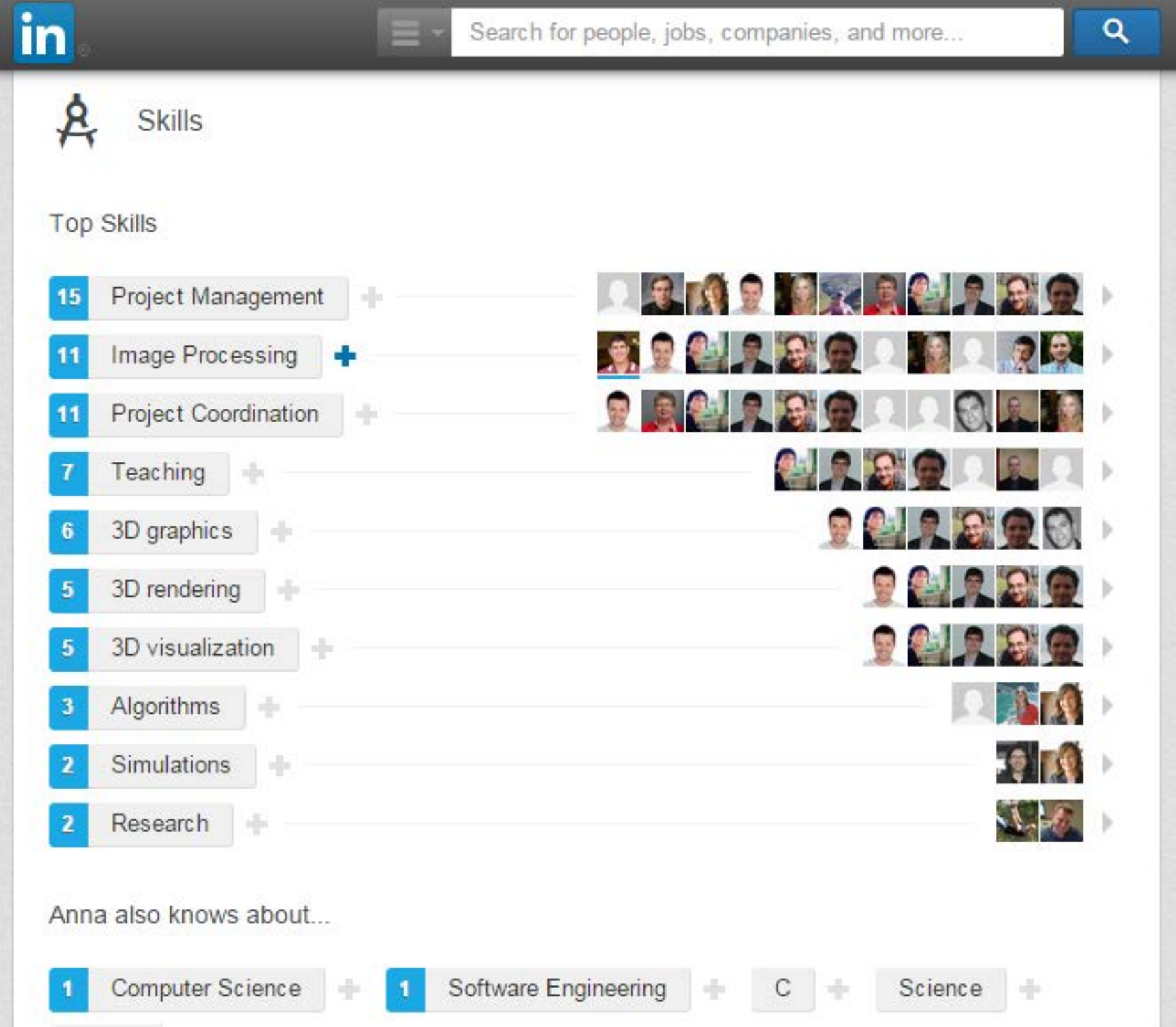}
	 \caption{Third endorsement mechanism: Anna's complete list of skills, which can be endorsed individually}
	 \label{fig:list}
\end{center}	
\end{figure}

We make a brief discussion about \LinkedIn's endorsement mechanism, which may be useful for the reader who is unfamiliar with the social network. The process starts when some user, say Anna, declares her skills (in this respect, \LinkedIn \ differs from other social networks, such as \ResearchGate, where users can suggest skills to their contacts). Then, Anna's contacts can endorse her for those alleged skills in three different ways: 

\begin{enumerate}
\item When one of Anna's contacts (say Ben) opens Anna's profile, the system presents Ben with a list of Anna's presumed skills, and asks Ben whether it is true that Anna possesses those skills. By pressing a single button Ben can endorse Anna for all the skills in the list. We may call that mechanism \emph{batch endorsement}. Figure \ref{fig:batch} shows the dialog box that is presented to Ben. The main advantage of batch endorsement is that it requires very little effort by Ben, since he only has to press a single button. Nevertheless, batch endorsement may be a source of inconsistencies, since the batch list presented to Ben is usually made up of several unrelated skills, and not necessarily those skills where Ben is an expert. It is true that Ben may remove some of the skills from the list, but that requires some additional effort. 
\item After Ben has batch-endorsed Anna, he is then asked to endorse other users, one skill at a time. The people appear in groups of four, and their order of appearance, as well as their skills, seem to be random. Figure \ref{fig:four} shows a group of four users waiting to be endorsed. This endorsement mechanism is more specific, but also more time-consuming than batch endorsement, and people usually skip it after a few clicks. 
\item Finally, if Ben wants to endorse Anna for some specific skill which did not appear, either in the initial batch list or in the subsequent endorsement suggestions, then he has to go to Anna's list of skills, and click on the specific skill he wants to endorse Anna for. This is by far the most reliable mechanism for endorsement, but it requires Ben's determination to make the endorsement. In Figure \ref{fig:list} we can see Anna's list of skills, which can be clicked on individually. 
\end{enumerate}

%--------------------
%--------------------

%--------------------
%--------------------

%--------------------
%--------------------

Now it becomes evident why there are inconsistencies in people's endorsements. In fact, we can say that practically all profiles surveyed by us contain some inconsistency, although some inconsistencies are more obvious than others, and the actual percentage may vary according to the definition of inconsistencies that is agreed upon. In any case, whatever the convention adopted, the percentage of profiles containing some form of inconsistency will be very close to 100$\%$. 

We have manually surveyed 100 profiles from a pool of 3250, mainly from the areas of Mathematics and Computer Science. Our sampling method consisted in a random DFS of depth two with backtracking. We started at the root profile (belonging to one of the authors), then  checked for inconsistencies, and selected one of the contacts at random, by generating a random number modulo the total number of contacts in the profile. Then we repeated the process from the new profile. Since the contacts are not visible for the profiles located at distance two (or greater) from the root profile, when we arrive at some profile of the second level, we check for inconsistencies and bactrack. Thus, our pool consists of all the profiles within distance two from the root profile. 

In the literature it is possible to find several sampling methods that are probably more effective than ours, but at this point, effectiveness and accuracy are not a concern, since we are not collecting any formal statistics. Our main purpose at this point is to find examples of inconsistencies, and classify the most significant inconsistencies encountered. Roughly speaking, the inconsistencies can be classified in the following categories: 

\begin{itemize}
\item Inconsistencies associated with the existence of hierarchies among skills. These can be subdivided into two groups:
\begin{itemize}
\item Bottom-up inconsistencies: A user is endorsed for some sub-category of a larger category, but not for the larger category. This is by far the most common inconsistency we have encountered. For example, some users have many endorsements for several programming languages, but do not have any endorsement, or have very few endorsements for the skill \lq Programming\rq \ itself, even though they have declared the skill \lq Programming\rq \ in their profiles. Also in relation to programming, some users have several endorsements for one or more object-oriented programming languages, such as Java, but are not endorsed for the skill \lq Object-Oriented Programming\rq. Still other users have been endorsed for \lq Object-Oriented Programming\rq, but not for \lq Programming\rq. In a different domain, some users have been endorsed for several mathematical skills, e.g. \lq Graph Theory\rq, \lq Discrete Mathematics\rq, \lq Applied Mathematics\rq, \lq Mathematical Modeling\rq, \lq Optimization\rq \ etc., but not for \lq Mathematics\rq. Finally, some users have been widely endorsed for \lq Digital Signal Processing\rq, \lq Digital Image Processing\rq, \lq Image Segmentation\rq, but not for the more generic \lq Image Processing\rq. 
\item Top-down inconsistencies: A user is endorsed for some category, but is not endorsed for any sub-category of the larger category. For example, some users have been endorsed for the skill \lq Programming\rq, but not for any specific programming language. 
\end{itemize} 
\item Inconsistencies associated with the existence of synonyms for a skill, or translations in different languages: A user is endorsed for some skill, but lacks endorsements in other skills that are synonyms of the first one. For example, several users have been endorsed either for the skill \lq Simulation\rq \ or for \lq Simulations\rq, but not both. Some users have been endorsed for some skill (say \lq Programming\rq) in a language other than English, but not in English. 
\item Inconsistencies between the information contained in the endorsements, and the information contained in the rest of the profile, or the public information available about the user. E.g. some user, who is a renowned expert in some area, is not endorsed for the corresponding skill. This may happen if the user himself has not declared the skill. For two concrete examples, Prof. Edy Tri Baskoro, who is a renowned graph theorist, is not endorsed for \lq Graph Theory\rq, and Prof. Cheryl Praeger, who is a renowned group theorist, is not endorsed for \lq Group Theory\rq.  
\end{itemize} 

This taxonomy does not attempt to cover all the situations encountered, which might be considered inconsistencies. It is very difficult to compile comprehensive statistics here, due to the huge diversity of cases, and to the subjectivism associated with defining inconsistent behaviour, but in any case it becomes quite clear that the endorsement mechanism offers some scope for improvement. 

\end{document}